\documentclass[aps,prb,twocolumn,superscriptaddress]{revtex4}
\usepackage{graphicx}
\usepackage{amssymb,amsmath}
\usepackage{subfigure}

\newcommand{\kt}{$k_{\rm B}T$}

\begin{document}
\bibliographystyle{apsrev}

\title{Self-assembly of artificial microtubules}
\author{Shengfeng Cheng}
\email{sncheng@sandia.gov}
\affiliation{Sandia National Laboratories, Albuquerque, NM 87185, USA}
\author{Ankush Aggarwal}
\affiliation{Department of Mechanical and Aerospace Engineering, 
University of California, Los Angeles, CA 90095, USA}
\author{Mark J. Stevens}
\affiliation{Sandia National Laboratories, Albuquerque, NM 87185, USA}

\date{\today}

\begin{abstract}
Understanding the complex self-assembly of biomacromolecules is
a major outstanding question.
Microtubules are one example of a biopolymer that possesses
characteristics quite distinct from standard synthetic polymers that are derived from
its hierarchical structure.
In order to understand how to design and build artificial polymers
that possess features similar to those of microtubules, we have
initially studied the self-assembly of model monomers into a
tubule geometry.
Our model monomer has a wedge shape with lateral and vertical binding sites
that are designed to form tubules.
We used molecular dynamics simulations to study
the assembly process for a range of binding site interaction strengths.
In addition to determining the optimal regime for obtaining tubules, we have
calculated a diagram of the structures that form over a wide range
of interaction strengths.
Unexpectedly, we find that the helical tubules form, 
even though the monomer geometry is designed for nonhelical tubules.
We present the detailed dynamics of the tubule self-assembly process and
show that the interaction strengths must be in a limited range to allow
rearrangement within clusters.
We extended previous theoretical methods to treat our
system and to calculate the boundaries between different structures in
the diagram.
\end{abstract}

\maketitle

\section{Introduction}

Self-assembly of macromolecules or nanoparticles into polymers or superlattices is an important route to produce
higher-level structures with distinct properties for numerous applications.\cite{MannNM09,science295}
Various assembled shapes can be formed, such as spheres, sheets and tubes, 
depending on the structural features encoded in building blocks,
their mutual interactions, and possible guiding from environment.
In biological systems, self-assembly of biomolecules into extended structures 
is very common and provides crucial functions. 
One example is the self-assembly
of microtubules (MTs) from $\alpha$- and $\beta$-tubulin heterodimers in cells. 
As key components of cytoskeleton,
MTs play crucial roles in cell structure and movement,
intracelluar protein transport, and cell division.\cite{Desai97,NogalesAR00} 
Besides being an important biological macromolecule,
MTs are special polymers that possess several properties distinct from
standard synthetic polymers, e.g. tubular structure, 
reconfigurable through depolymerization-polymerization cycles, and the substrate for motor proteins.
MT assembly is a very complex process and thus difficult to characterize.
Understanding the assembly of macromolecules 
that possess a subset of these properties would be remarkable and is our
more practical goal.
Our group is interested in developing polymers that are ``artificial" MTs, 
i.e. possess many characteristics of MTs, 
but are assembled from different monomers and constituent molecules.\cite{McElhanonMM10}
In this paper we present results of an initial simulation study of the assembly of artificial MTs.
We focus on wedge-shaped monomers designed to self-assemble into tubular structures upon mutual binding
and investigate how the assembled structures depend on binding strengths.

Tubulin monomers that make up MTs are about 4 nm in size.\cite{NogalesNature98}
They self-assemble into a helical tubular structure when their 
concentration is above certain threshold or when there exist 
appropriate nucleation sites, such as $\gamma$-tubulin ring complex 
in eucaryotic cells.\cite{KollmanNRMCB11,ZhengNature95}. 
Each helical turn typically contains 13 tubulin dimers {\it in vivo},
but the number can vary from 11 to 15 {\it in vitro}.\cite{Desai97}
The diameter of MTs is around 25 nm, and their length can be as long as 25 $\mu$m.

Another type of biological macromolecules that can self-assemble into tubes are 
surface layer (S-layer) proteins, which make up the cell wall of 
prokaryotic organisms (bacteria and archaea).
S-layer proteins typically crystallize into two-dimensional arrays
with various symmetries.\cite{Pum95,Jarosch01} 
However, particular S-layer proteins under the right conditions
can form open tubes with the diameter being controlled
by adding or removing amino acids.\cite{Mertig01,BobethLangmuir11,KorkmazNT11a}
Various assemblies of S-layer proteins 
can serve as templates to grow other extended nanostructures
such as superlattices of metallic clusters.\cite{Mertig01,Sleytr07,KorkmazNT11b}

Tubular structures are also found to form in self-assembly
of macromolecules that are usually amphiphilic.
\cite{science295,Schnur93,Orr99,Bong01,Yan04,
Hill04,Shimizu05,Shimizu08,Zhang07,Kwak10,Hamley11,Coleman11} 
The location of hydrophobic and hydrophilic groups and other non-covalent interaction sites 
results in the formation of tubes. 
The key issue is to understand what features (shapes, chirality, binding strengths, etc.) 
of the molecular building blocks under certain formation conditions
are required to yield tubes with desired structures (e.g., the tube diameter) 
and physical properties.
However, there are so many variables that it becomes difficult
to explore the whole phase space experimentally.
This is where molecular simulations might be most useful 
to identify the range of parameters (particularly binding strengths) 
most appropriate to form tubes with controlled structures.

An important, related example of self-assembly that for which some important aspects have been 
determined is the formation of a viral capsid from just one protein.\cite{Caspar62}
Recent theoretical and simulation work have addressed the nature of this self-assembly
and revealed some simplifying aspects of the 
assembly.\cite{BruinsmaPRL03,ZandiPNAS04,ZandiBJ06,RapaportPRL08,Roos10} 
The assembly of the protein monomer into any possible structure, 
not just into the capsid with icosahedral symmetry, has been examined.
While in the continuum limit, the free energy minimum favors icosahedral symmetry, for smaller capsids
there are structural parameters that drive the specific geometry.\cite{BruinsmaPRL03}
Rapaport used MD simulations to study the formation of viral capsids from 
capsomers modeled as truncated pyramids with interaction sites 
on their lateral surfaces.\cite{RapaportPRL08,Rapaport10}
Besides showing that these capsomers are able to self-assemble into capsids, 
his work found important general design aspects of the
capsomer for assembly to occur.
He found that there is a narrow range of interaction strengths for the assembly of capsids.
In addition, the simulations showed that reversibility along the assembly pathway is crucial for 
efficient production of complete shells by suppressing kinetic traps. 

Similar to Rapaport,\cite{RapaportPRL08} 
we use a 3D monomer with a wedge shape that promotes assembly into tubes. 
Reproducing all the properties of MTs is out of the question and 
given our interest in artificial systems, not our main goal.
One initial purpose is to build a minimal model of the self-assembly of tubular structures. 
From the viral capsid results, we expect there to be a narrow range of 
interactions that will yield tubule self-assembly.
Determining this range is essential before any further, more complex
issues can be addressed.
Our wedge monomers have lateral and vertical (also called ``longitudinal'' in MT literature) 
binding sites on their surfaces, whose interaction strengths are independently varied. 
We have thus determined the optimal range of interaction strengths between monomers 
that achieve tubules 
and studied other structures that form more generally at other interaction strengths.
We observed a rich set of assembled structures reminiscent of assembly 
of the coat protein into the tobacco mosaic virus.\cite{Klug99}

The assembly of wedge-shaped monomers is in the class of
associating fluids and involves living polymerization 
under certain circumstances.
In living polymerization, polymer chain can grow and shrink; ultimately
there is an equilibrium between single monomers and a distribution of polymer
lengths.
The systems of associating fluids or those undergoing living polymerization can be modeled
as an ensemble of particles allowing monomers to
associate to form dimers, trimers, and long chains or clusters.
The theoretical description of the assembly of these systems 
usually starts from either Wertheim theory,\cite{Wertheim84,SciortinoJCP07} 
which is essentially a perturbation theory, or Flory-Huggins theory.\cite{DudowiczJCP99}
Our model wedge, particularly when only two opposite sides are active, is in
the class of polymer systems.
In this paper we have developed a Flory-Huggins type theory that particularly treats 
the features of our model to calculate the structure diagram 
as a function of the interaction parameters.

The remainder of this paper is organized as follows. 
In Sec.~II, the simulation methods are described.
Then in Sec.~III a thermodynamic theory, 
along the line of Flory-Huggins lattice model of polymerization,
is developed for the self-assembly of anisotropic objects, which
will be used later to explain the assembly behavior 
observed in MD simulations.
The MD results are presented in Sec.~IV.
Finally, discussion and conclusions are included in Sec.~V and VI.

\section{Simulation Methods}
\label{sec:method}

\begin{figure}[ht]
  \subfigure[]{
    \begin{minipage}[b]{0.14\textwidth}
      \centering
      \includegraphics[width=1in]{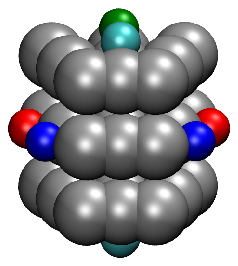}
     \end{minipage}}%
  \subfigure[]{
    \begin{minipage}[b]{0.16\textwidth}
      \centering
      \includegraphics[width=0.9in]{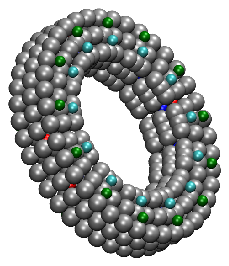}
     \end{minipage}}%
  \subfigure[]{
    \begin{minipage}[b]{0.15\textwidth}
      \centering
      \includegraphics[width=1in]{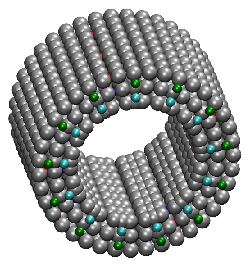}
     \end{minipage}}
\caption{(a) A wedge-shaped monomer; 
(b) An ideal ring formed by 13 monomers;
(c) An ideal tube with 3 rings stacked.
Each monomer consists of 35 sites, of which 27 particles (gray) 
form its back-bone and the other 8 sites on the surfaces (in color) 
are the locations of the attractive interaction centers.
Attractions only act between
sites with the same color.}
\label{figure:geometry}
\end{figure}

In order to have a monomer that will self-assemble into a tubule structure,
the monomer is chosen to have a truncated wedge shape (Fig. \ref{figure:geometry}).
To make a wedge, we started with 27 particles in $3\times 3\times 3$ simple cube.
The distance between two neighboring sites is $1\sigma$ 
in the initial cube.
The cube is deformed into a wedge shape such that 
13 of them will join to form a closed ring. 
The back layer is unchanged, while the front and middle layers are compressed 
such that the angle
made by the two sides fits 13 monomers in a ring.
The particles (gray spheres in Fig. \ref{figure:geometry}) interact 
purely repulsively through the Lennard-Jones (LJ) potential
\begin{equation}
U(r)=4\epsilon\left[ (\sigma/r)^{12}-(\sigma/r)^6
-(\sigma/r_c)^{12}+(\sigma/r_c)^6 \right],
\label{LJPotential}
\end{equation}
where $r$ is the distance between the center of two particles, 
and $\epsilon$ is the unit of energy. 
The potential is truncated at $r_c=1.0\sigma$ 
to make the interaction purely repulsive, which indicates that two monomers will 
repel each other when they get very close.

The attractive component of the monomer-monomer interaction occurs at sites located on
the wedge surface, specifically on pairs of sites on opposite faces (colored spheres in
Fig. \ref{figure:geometry}).
To control the orientation of two monomers that bind,
two distinct attractive sites are necessary on a surface
in order to break the symmetry of the surface-surface interaction.
In Fig. \ref{figure:geometry} sites with the same color are attractive and sites with 
different colors have no interaction (effectively repulsive due to the other interactions).
The red and blue sites on the lateral sides promote the formation of rings.
In order for the rings to stack to form a tubular structure, the two vertical 
(top and bottom) faces also have a pair of attractive interaction sites (cyan, green).
The attractive interaction is given by
\begin{equation}
U(r)=-A\left[1+{\rm cos}\left(\frac{\pi r}{r_a}\right)\right],
\label{SoftPotential}
\end{equation}
where $A$ is the strength of the potential and the range is given by $r_a$.
With this potential the strength and range of the attraction can be
independently adjusted.
In this work, we will only use $r_a = 1\sigma$.
The value of $A$ for the lateral and vertical surfaces will be independently
varied.
We will denote the potential strengths for the lateral surfaces to be $A_L$
and for the vertical surfaces to be $A_V$. 

The dynamics of self-assembly was obtained by performing MD
simulations using the LAMMPS simulation package.\cite{plimpton95,lammps}
Each wedge molecule is treated as a rigid body.
The equations of motion were integrated using a velocity-Verlet algorithm 
with a time step $\delta t =0.005\tau$,
where $\tau=\sigma(m/\epsilon)^{1/2}$ is the unit of time
and $m$ is the mass of one particle.
The simulations to calculate the structure diagram were run for 
$5\times 10^5$ to $2.5\times 10^6\tau$.
Unless otherwise noted, the simulations are for systems composed of $1000$ wedges.
To create an initial state without overlap of wedge monomers, the positions
of the wedge monomers was taken from an equilibrated LJ system at 
density $0.2\sigma^{-3}$. 
Each LJ particle was replaced by a randomly oriented wedge particle
and the system was scaled by the wedge particle size resulting
in the volume fraction of wedges being $3.85\%$.
The temperature of the system is kept at $1.0\epsilon/k_{\rm B}$,
where $k_{\rm B}$ is the Boltzmann constant, with
a Langevin thermostat of damping rate $1.0\tau^{-1}$.
Since the important energy scale is $k_{\rm B}T$ and since
$\epsilon = k_{\rm B}T$, we will use $k_{\rm B}T$ as the energy unit.

\section{Thermodynamic Theory of Artificial Microtubule Assembly}
\label{sec:theory}

To help understand the simulation results, we adapt
a Flory-Huggins theory to describe
the assembly behavior of wedge-shaped monomers.\cite{BruinsmaPRL03,ZandiBJ06} 
Suppose the simulation box is divided into a three dimensional lattice 
with $M$ cells, each of which can only accommodate
at most one monomer. 
The entropy of the system is clearly
determined by the number of ways to put $N$ monomers 
into these $M$ cells. 
As introduced in Sec.~\ref{sec:method}, 
monomers simulated in this paper
can form two types of bonds, lateral and vertical.
The lateral bonding drives monomers to form closed rings, while
the vertical bonding leads to the formation of filaments. 
Ideally, the appropriate combination of these two types of bonds yields tubes.

To get the critical bonding strength required for self-assembly,
we first consider the simple case where monomers assemble into linear
chains (e.g. $A_L=0$). 
Treating the linear case first allows us to follow much of the previous 
theoretical works.\cite{OosawaJMB62,DudowiczJCP99,ZandiBJ06}
Denote the number of $p$-segment chains 
(consisting of $p$ monomers) as $n_p$ and
assume the maximum value of $p$ is $p_{\rm max} \le N$,
the first constraint is from the conservation of total monomers and reads
\begin{equation}
N = \sum_{p=1}^{p_{\rm max}} p n_p.
\label{EqNumConstraint}
\end{equation}

The Helmholtz free energy of our system can be written as 
(see Appendix for a detailed derivation)
\begin{equation}
\begin{split}
F =\frac{1}{\beta}
\sum_{p=1}^{p_{\rm max}}n_p\left( \beta (p-1) \Delta g - \ln z +  \ln \frac{n_p}{M} - 1\right),
\end{split}
\label{EqFreeE}
\end{equation}
where $(p-1)\Delta g$ is the total energy gain to form a $p$-segment chain,
which has $p-1$ bonds, $z$ is the coordination number of the lattice,
and $\beta = 1/k_{\rm B}T$.
This free energy has a form very similar to that derived by Zandi {\it et al.}
for viral capsid assembly.\cite{ZandiBJ06}
Minimizing the above free energy with the constraint Eq.~(\ref{EqNumConstraint})
leads to the distribution of chains as
\begin{equation}
n_p = zM \exp (\beta \Delta g) \exp \left[-\beta p \left(\Delta g -\mu\right)\right],
\label{EqChainDist}
\end{equation}
where $\mu$ is a Lagrange multiplier and 
has a physical meaning of the chemical potential.
Note that the law of mass action is implicit in the above expression of $n_p$.
Since $N/M$ is the volume ratio of all monomers, $\mu$ can be 
determined by combining the distribution $n_p$ and Eq.~(\ref{EqNumConstraint}).
Let $x= \exp\left[-\beta \left(\Delta g -\mu\right)\right]$,
we obtain
\begin{equation}
z \exp\left(\beta \Delta g\right)\sum_{p=1}^{p_{\rm max}} px^p = \frac{N}{M},
\end{equation}
Complete the summation and take the limit $p_{\rm max}\rightarrow \infty$, 
also notice that $x<1$ to be physically meaningful, 
we arrive at the equation
\begin{equation}
z \exp\left(\beta \Delta g\right)\frac{x}{(x-1)^2}= \frac{N}{M},
\label{EqAssemCond}
\end{equation}
from which $x$ can be computed from known $N/M$ and $\Delta g$.
Then the chemical potential $\mu$ is given by 
$\mu = \Delta g + \beta^{-1} \ln x$.
With $\mu$ determined, the distribution of chains $n_p$
is easily computed.
If we identify $z^{-1} \exp (-\beta \Delta g)$ 
as the equilibrium constant of dimerization, 
Eq.~(\ref{EqAssemCond}) becomes identical to Eq.~(5) in Ref.~40,
which was derived directly from the law of mass action.

The assembly is purely controlled by the bonding energy $\Delta g$
when the total number of starting monomers and the volume of the box
are fixed. At small $\Delta g$, monomers dominate and $n_1$
is much larger than all $n_p$'s with $p \ge 2$.
With the increase of the magnitude of $\Delta g$ ($ < 0$),
$n_1$ monotonically decreases.
The onset of assembly is identified with the condition
that half of the all starting monomers are in the assembled state, 
i.e., in $p$-segment chains with $p \ge 2$,
while the remaining half is in the form of free monomers.\cite{BruinsmaPRL03}
Using Eq. (\ref{EqChainDist}) and the condition $n_1=N/2$ we obtain
\begin{equation}
2z{\rm e}^{\beta \mu} = \frac{N}{M},
\end{equation}
since the number of leftover monomers is $n_1=zM{\rm e}^{\beta \mu}$.
Combining this equation with Eq.~(\ref{EqAssemCond}) yields
the critical bonding strength 
\begin{equation}
\beta \Delta g_{cb} =  \ln \frac{N}{M} -  \ln z -  \ln (2-\sqrt{2}).
\end{equation}
For our simulations, $N/M=0.0385$, so
$\Delta g_{cb} = -4.5~k_{\rm B}T$ if we take $z=6$.

The theory also allows us to estimate the bonding strength required
for the mass formation of $p$-segment chains. 
When $\left|\Delta g \right|$ increases from 0,
$n_2$ starts to grow, eventually surpasses $n_1$,
and becomes the dominant mode.
Then $n_2$ decreases and gets passed by $n_3$, and likewise for $n_p$ with larger $p$.
Therefore $n_p$ ($p \ge 2$) shows a maximum at certain $\Delta g_c (p)$
that depends on $p$.
We identify this $\Delta g_c (p)$ as the critical bonding strength that
signals the assembly transition to $p$-segment chains.
The corresponding equation is
\begin{equation}
\frac{\partial n_p}{\partial \Delta g} = 0.
\label{phasecriterion}
\end{equation} 
Combining this condition with Eqs.~(\ref{EqChainDist}) 
and (\ref{EqAssemCond}) yields
\begin{equation}
x = \frac{p-1}{p+1},
\end{equation}
and a simple expression for the critical bonding energy
\begin{equation}
\beta \Delta g_c(p) =  \ln \frac{N}{M} -  \ln z -  \ln \frac{p^2-1}{4}.
\end{equation}
Note that $\left|\Delta g_c(2) \right| = 4.76~k_{\rm B}T$
is larger than $\left|\Delta g_{cb} \right|$
that signals the overall assembly transition.
Also note that $\Delta g_c(p) \rightarrow -\infty$ when $p\rightarrow 1$,
which seems counterintuitive but is indeed 
the byproduct of the approximate $F$ in Eq.~(\ref{EqFreeE}) from the lattice model.

The above theory is developed for the straight polymerization
of wedge monomers through vertical bonding.
Our simulations further indicate that 
it also works in the case of ring formation,
i.e., the case where $A_L \neq 0$ and $A_V = 0$.
The main difference is that a ring only contains 
certain number of monomers. 
In this case, $p_{max}$ is finite.
However, it turns out that in the range of $A_L$ where
rings do form, the results from a finite $p_{max}$, 
such as $p_{max}=13$ for ideal rings, and those
from an infinite $p_{\max}$ are very close.
The reason is that chains with $p > 13$
contribute little to the summation in Eq.~(\ref{EqChainDist}),
which is the constraint used to determine $n_p$.

To further describe the formation of tubes, 
we need to take into account simultaneously both
the vertical (chain formation) and lateral (ring formation) bonding interactions.
It turns out a simple picture can reasonably 
explain our simulation results as presented in the next section.
The basic idea is that as long as both the vertical and lateral bonding
can occur and stay stable (its meaning will be clarified in the next section), 
then it is possible for monomers to self-assemble into tubes,
through a cluster growth process.
For example, for a system dominated by $p$-segment 
linear chains formed through vertical bonding, 
the tubes will emerge as long as the side-to-side (lateral) bonding 
of these chains can occur and stay stable.
In this case, the required lateral bonding strength will depend on $p$,
in addition to the $p$-dependence of the vertical bonding $\Delta g_c(p)$ 
that controls the transition to $p$-segment chains.
With a mapping between $\Delta g$ and $A$, 
this will lead to a $(A_V,A_L)$ curve that signals the transition to 
the tubular structure phase from the filament phase.
In another example, if $A_L$ is large enough to induce rings, then
tubes will appear when $A_V$ is strong enough to lead to stable bonding
that stacks rings together.
As shown later, these arguments lead to accurate estimates of critical bonding strengths
that are required for the tube formation
and they agree well with our MD calculations.

\section{Results}

\subsection{Structure Diagram}

We have varied the interaction strengths $A_L$ and $A_V$ primarily to determine the 
appropriate values to form tubules.
Given the experience with capsid assembly,\cite{RapaportPRL08}
we expect that there will be a narrow range of interaction strengths that are good for
forming tubules without defects.
We are also interested in the competition between the lateral and vertical interactions.
In natural MTs, it was viewed for a long time 
that long protofilaments formed and merged to 
form the tubule.\cite{ChretienJCB95,Desai97,GardnerCurrOp08}
In other words, the vertical interaction was stronger and vertical growth dominated.
More recent experiments find that the growing end of a MT 
can vary by lengths corresponding to only a few tubulin dimers 
or even just one dimer, 
which implies that the vertical and lateral growth 
and interaction strengths are not 
so distinct.\cite{KerssemakersNature06,SchekCB07,MozziconacciPLoS08}
In cells, MTs typically grow from a preexisting nucleation center,
e.g., the ring complex involving $\gamma$-tubulin.\cite{KollmanNRMCB11,ZhengNature95}
There are no such preformed nucleation sites in our simulations. 
However, small clusters formed at an earlier stage of the assembly
serves such a role as long as they can stay stable for a 
substantial interval. During the assembly, such clusters capture
free monomers or even other clusters to grow into tubes or
larger clusters. More details will be included
in the later subsection on assembly kinetics.

\begin{figure}[t]
\centering
\includegraphics[width=2.75in]{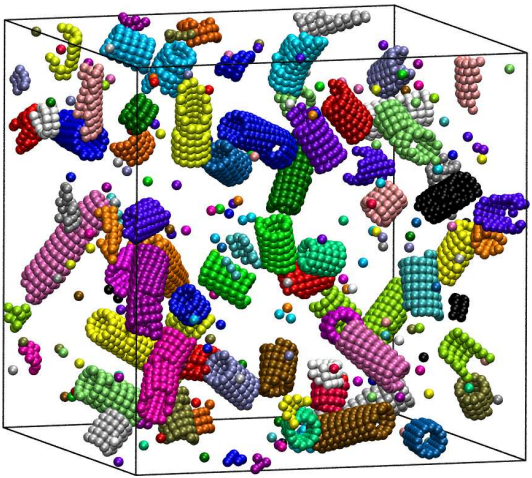}
\caption{Image of system with $A_L=4.4~k_{\rm B}T$ and $A_V=2.6~k_{\rm B}T$
showing many tubules formed.
The wedge monomers are represented by spheres. 
Monomers in the same cluster have the same color.
There are only 32 colors; some colors are used more than once.
}
\label{figure:tubules}
\end{figure}

The wedge monomers do form tubules as shown in Fig.~\ref{figure:tubules}.
This image shows the system at $A_L=4.4~k_{\rm B}T$ and $A_V=2.6~k_{\rm B}T$,
which is one of the parameter sets for which tubule structures readily form.
This system is a large one that has 5000 monomers and many, long tubules
are seen to have formed.
One unexpected result is that there are multiple tubules that are helical.
For example, the pink tubule near the lower front of the image clearly shows 
the helical turn of monomers instead of the straight stacking of rings.
The helical turn can also be seen clearly in the cyan tubule just above and to the right of the pink tubule.

\begin{figure}[htb]
\centering
\includegraphics[width=3in]{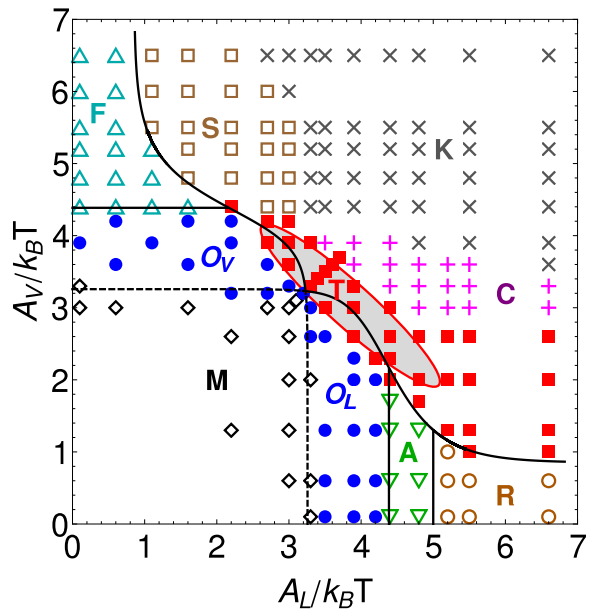}
\caption{The structure diagram of the assembly of 
the wedge-shaped monomers from MD simulations (symbols)
and comparison with the prediction of the thermodynamic theory (lines):
monomers (M) (open diamonds), 
oligomers (${\rm O_L}$ and ${\rm O_V}$) (solid circles),
partial rings or arcs (A) (downward open triangles), 
full rings (R) (open circles),
full tubules (T) (solid squares), 
filaments (F) (upward open triangles),
large sheets (S) (open squares),
multiple clusters (C) (pluses),
kinetically trapped percolated cluster (K) (crosses).}
\label{figure:phase}
\end{figure}

\begin{figure*}[tb]
  \subfigure[]{
    \begin{minipage}[b]{0.25\textwidth}
      \centering
      \includegraphics[width=1.5in]{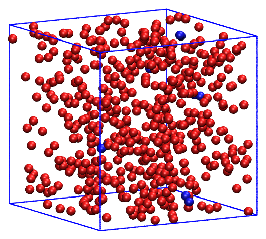}
     \end{minipage}}%
  \subfigure[]{
    \begin{minipage}[b]{0.25\textwidth}
      \centering
      \includegraphics[width=1.5in]{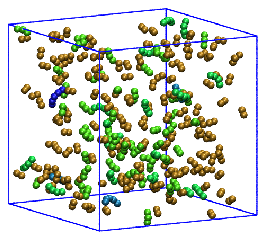}
     \end{minipage}}%
  \subfigure[]{
    \begin{minipage}[b]{0.25\textwidth}
      \centering
      \includegraphics[width=1.5in]{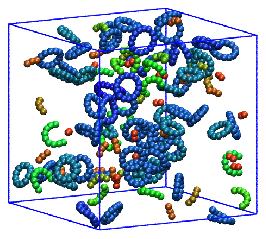}
     \end{minipage}}%
  \subfigure[]{
    \begin{minipage}[b]{0.25\textwidth}
      \centering
      \includegraphics[width=1.5in]{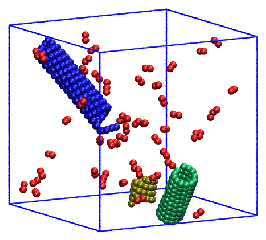}
     \end{minipage}}\\
  \subfigure[]{
    \begin{minipage}[b]{0.25\textwidth}
      \centering
      \includegraphics[width=1.6in]{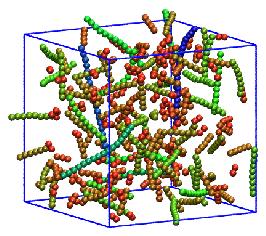}
     \end{minipage}}%
  \subfigure[]{
    \begin{minipage}[b]{0.25\textwidth}
      \centering
      \includegraphics[width=1.5in]{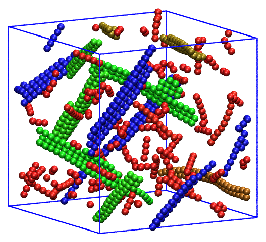}
     \end{minipage}}%
  \subfigure[]{
    \begin{minipage}[b]{0.25\textwidth}
      \centering
      \includegraphics[width=1.5in]{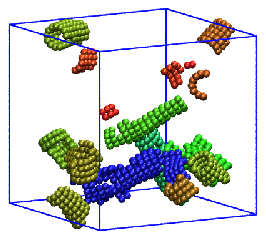}
     \end{minipage}}%
  \subfigure[]{
    \begin{minipage}[b]{0.25\textwidth}
      \centering
      \includegraphics[width=1.5in]{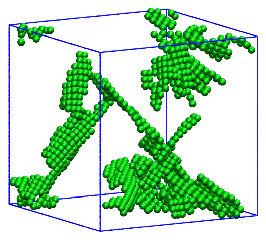}
     \end{minipage}}
\caption{Images of states in structure diagram:
(a) the monomer state (M state) at $A_L=A_V=3.0~k_{\rm B}T$;
(b) partial rings and arcs (A state) at $A_L=4.4~k_{\rm B}T$ and $A_V=1.7~k_{\rm B}T$;
(c) rings (R state) at $A_L=5.5~k_{\rm B}T$ and $A_V=0.6~k_{\rm B}T$;
(d) assembly of tubes (T state) at $A_L=3.9~k_{\rm B}T$ and $A_V=2.6~k_{\rm B}T$;
(e) filaments (F state) at $A_L=1.1~k_{\rm B}T$ and $A_V=5.2~k_{\rm B}T$;
(f) large sheets (S state) at $A_L=1.6~k_{\rm B}T$ and $A_V=5.2~k_{\rm B}T$.
(g) multiple clusters (C state) at $A_L=4.4~k_{\rm B}T$ and $A_V=3.3~k_{\rm B}T$.
(h) kinetically trapped percolated cluster state 
(K state) at $A_L=4.4~k_{\rm B}T$ and $A_V=5.2~k_{\rm B}T$.
For clarity, each monomer is represented as a sphere, and 
free monomers are only shown except in the panel (a).
Monomers are colored based on the size of the cluster they reside in (red to blue).
Note that periodic boundary conditions are in effect in all three directions 
so that monomers/clusters on opposite sides of the box might actually
belong to the same cluster.
}
\label{figure:eightstates}
\end{figure*}

Overall, the MD simulations find an interesting set of different structures as shown in
Fig.~\ref{figure:phase}.
Images of the corresponding structures are shown in Fig.~\ref{figure:eightstates}.
The images are colored based on cluster size.
Two wedge monomers are considered to be in the same cluster
if their mutual bonding energy is larger than $50\%$ 
of the full bonding energy $4A$.
We have analyzed the bonding energy distributions and 
this criterion has accurately captured all bonds 
and does not introduce any artificial ones.

In the lower-left region of the structure diagram the interaction strengths
are too small for assembly to occur.
The majority of the system is in the single monomer state.
A snapshot of such a system is given in Fig.~\ref{figure:eightstates}(a) (M state).
Along the $A_L=0$ line, standard polymerization of a straight living polymer
occurs, once the value of $A_V$ is large enough to yield stable bonds.
For $A_V \gtrsim 3.2$ \kt, stable oligomers start to emerge but single monomers still dominate.
We denote this state as $O_V$ but consider it only as a subset of the M state.
At $A_V = 4.4$ \kt{} the number of single monomers decreases to $50\%$
of the total number of the starting monomers.
Similarly, along the $A_V=0$ line, oligomers appear when $A_L \gtrsim 3.2$ \kt{}
and at $A_L > 4.4$ \kt{} the monomer count is below $50\%$.
Between $A_L = 4.4$ \kt{} and 5 \kt,
the polymers are partial rings or arcs 
(A state, see Fig.~\ref{figure:eightstates}(b); 
the interaction strength is not large enough for full rings.
Beyond $A_L = 5 $ \kt, individual rings form 
(R state, see Fig.~\ref{figure:eightstates}(c)).

Moving above the $A_V = 0$ line for $A_L \gtrsim 3.0$ \kt, the structure does not change until a large enough
value is reached to achieve vertical assembly.
For $A_L$ values that have produced even partial rings, the combination of
two partial rings occurs at low values of $A_V$, since the multiple monomers
bind between the partial rings yielding the extra energy needed to form a larger cluster.
Full tubules occur in the region denoted with solid squares in Fig. \ref{figure:phase} 
(T state)
and an example is shown in Fig. \ref{figure:eightstates}(d).
With respect to the ease of assembly, 
we find that tubules are more well formed (i.e. fewer defects) 
in the region circled by the ellipse in the figure, where $A_L \gtrsim A_V$.
The long tubules (blue and green) in Fig. \ref{figure:eightstates}(d) are helical, which was not
expected given that the wedge was designed to form nonhelical tubes.

At relatively small $A_L$, in the region where $A_V\gtrsim 4~k_{\rm B}T$, 
oligomers and long filaments form as vertical bonding dominates.
One example of this filament state (labeled F) is shown in 
Fig. \ref{figure:eightstates}(e).
When $A_L\gtrsim 1~k_{\rm B}T$, filaments can bond side-by-side and 
curved sheets form (S state, see Fig. \ref{figure:eightstates}(f)). 
Once the value of $A_V > 5 \sim 6 ~k_{\rm B}T$, sheets grow very long 
due to very strong vertical bonding, and while
rings of monomers do form within some of the curved sheets as long as $A_L\gtrsim 2~k_{\rm B}T$, 
well formed tubules do not appear.
At even larger values of both $A_L$ and $A_V$, 
the system tends to get kinetically trapped.
Most or even all monomers join a dominating cluster that percolates the whole simulation box
(K state, see Fig. \ref{figure:eightstates}(h));
a diffusion limited aggregation is most likely occurring in this region.
The K state is identified by the criterion that the largest cluster contains
more than $50\%$ of the all monomers.
When $A_L$ is kept large and $A_V$ is lowered to around $4~k_{\rm B}T$,
systems are still in the kinetically trapped phase. However, now there exist
multiple smaller clusters in the system (C state, see Fig. \ref{figure:eightstates}(g)). 
These clusters are very stable and have many defects 
induced by the strong lateral bonding.
It is very difficult for them to fit to each other and fuse into tubes.

It must be emphasized that there are no sharp boundaries amongst
the S, K, C, and T states. In the cross-over region between these states,
very often various structures can coexist with each other. 
For example, in the C state, there are several clusters,
inside which closed rings may appear and there may 
also exist tubes. However, these tubes usually contain a lot of defects,
which can stay for a long time.
Furthermore, the distinction between the C state and the K state is just that 
in the K state the dynamics results in clusters 
that form early in time, and then collide and merge into a percolated structure.

Large $A_V$ does not favor the formation of tubes because 
long filaments form first, and when they initially bind,
they are not aligned parallel.
While they may in time align parallel, the filaments tend to be offset,
and the process for two filaments to diffuse to remove the offset is very
slow at best and not observed in the simulation.
In this case, even though the tube phase has the lowest free energy 
and is the thermodynamically equilibrium state, 
it is not achieved at least on the MD timescale. 

\begin{figure}[bt]
\centering
\includegraphics[width=2.5in]{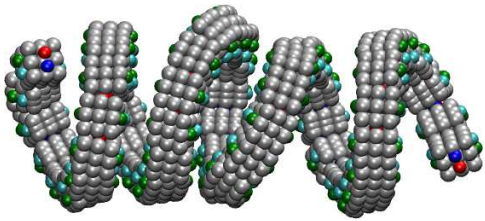}
\caption{A long helical twist formed at $A_L=8.8$ k$_B$T and $A_V=0$.}
\label{figure:twist}
\end{figure}

We also observed that when $A_L$ is very large and $A_V$ is small, 
there emerges interesting structures other than rings.
One example is shown in Fig.~\ref{figure:twist}, 
which is a long helical twist formed at
$A_L=8.8~k_{\rm B}T$ and $A_V=0$.
The number of monomers in this kind of twist can be much larger than $13$.
Furthermore, at a large $A_L$ where rings can be formed, 
the number of monomers in a turn varies substantially.
For example, at $A_L=5.5~k_{\rm B}T$ and $A_V=0.1~k_{\rm B}T$, 
we have found rings
containing 12 to 14 monomers with a peak at 13.
At $A_L=6.6~k_{\rm B}T$ and $A_V=0.1~k_{\rm B}T$ 
the number can vary anywhere from 11 to 16.
The rings with a large number of monomers usually have
a slightly twisted shape out of a plane 
because of the geometrical constraint of the wedge size and shape.
In these cases, tubes do form when $A_V$ is increased. 
However, the tubes tend to have many defects,
partly because the strong bonding interactions 
make the relaxation very difficult, 
which hinders the removal of structural defects through 
adjustment of bonds. 

\subsection{Comparison with Thermodynamic Theory}

A basic issue is the minimal well depth ($A$) necessary to achieve assembly. 
The structure diagram (Fig.~\ref{figure:phase}) shows that assembly does not occur
until $A_{L,V}$ is greater than about $3~k_{\rm B}T$.
Given that the monomers are rigid bodies with 6 total degrees of
freedom, the entropy per monomer is $3~k_{\rm B}T$.
Thus, we do not expect dimer formation until beyond this range of $A$.
How large a magnitude of $A$ must be to form dimers that
are stable enough to grow into trimers is an open question, 
which we have addressed for our system.
To determine the binding energy to form dimers 
we examined in more detail the interaction between a pair of wedges.
These dimer calculations provide a mapping between the model
parameters $A_{L,V}$ and the bonding energy $\Delta g$ 
in the thermodynamic theory outlined in Sec.~\ref{sec:theory}.
At the level of dimer formation, simulations show there is no essential difference 
between lateral and vertical bonding.
We directly calculated the bonding energy between two monomers 
starting in a bound dimer state, 
bonded either laterally or vertically, respectively.
Below we use $A$ to designate either $A_L$ or $A_V$.
Simulation run times were from $5\times 10^3\tau$ to $5\times 10^4\tau$; 
the longer run times were for cases near or beyond 
the stability limit where the bond had longer lifetimes.

\begin{figure}[ht]
\centering
\includegraphics[width=2.5in]{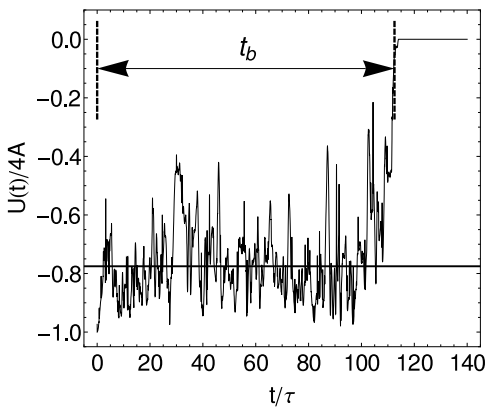}
\caption{The bonding energy $U(t)$ as a function of time for $A=3~k_{\rm B}T$. 
The horizontal line shows the mean bonding energy $U_B$ before 
the dissociation,
and the time $t_b$ is the bond lifetime in this MD run.}
\label{figure:BondEnergy}
\end{figure}

The instantaneous bonding energy of the dimer is denoted as $U(t)$
to indicate its temporal dependence. 
A typical example of $U(t)$ is shown in Fig.~\ref{figure:BondEnergy}
for $A=3~k_{\rm B}T$.
The debonding transition to the unbound state is quite sharp, which
allows us to define a bonding time $t_b$ that
describes the life time of the bond.
The mean bonding energy from the plateau of $U(t)$
at $t<t_b$ is defined as the energy per bond $U_B$,
which is shown as the horizontal line in Fig.~\ref{figure:BondEnergy}.

\begin{figure}[htb]
\centering
\includegraphics[width=2.5in]{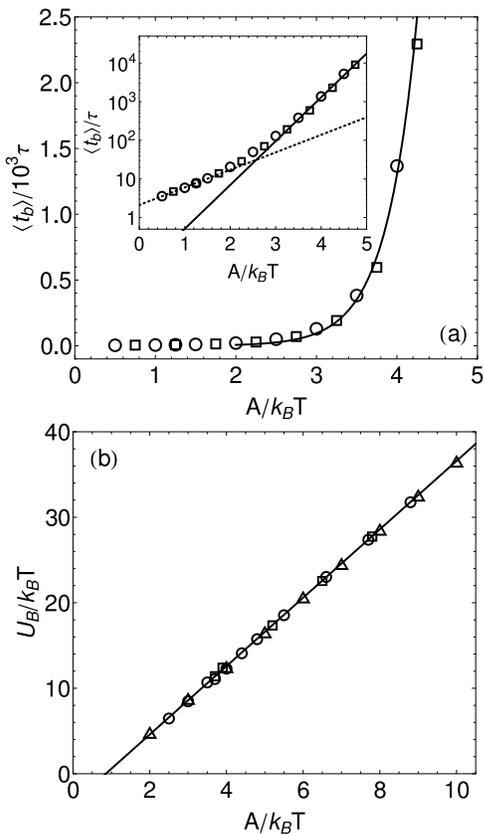}
\caption{(a) The mean bonding time $\langle t_b \rangle$ vs. $A$. 
Data are calculated with
dimers starting at full bonding either laterally (circles)
or vertically (squares). The inset
is the linear-log plot of the same data set in the main panel. 
The dotted (solid) line in the inset is the linear fit
to the first (last) 6 data points.
The solid line in the main panel is the corresponding exponential curve
for the last 6 data points.
(b) The bonding potential $U_B$ vs. $A$ for one bond calculated with
various starting configurations and temperatures: 
two monomers bonding laterally at $T=1.0\epsilon/k_{\rm B}$ (circles);
two monomers bonding vertically at $T=1.0\epsilon/k_{\rm B}$ (squares);
two monomers bonding laterally at $T=0.1\epsilon/k_{\rm B}$ (triangles).
Both $U_B$ and $A$ are normalized by $k_{\rm B}T$.
The solid line shows a linear fit $U_B = 4*A-3.41~k_{\rm B}T$.
}
\label{figure:mapping}
\end{figure}

By starting with the same initial dimer configuration 
and following different paths in the phase space (i.e. varying random
number seed in Langevin thermostat),
we have calculated the mean bonding time $\langle t_b \rangle$ as a function of
the soft potential strength $A$. 
Results on $\langle t_b \rangle$ vs. $A$ are shown in the main panel of
Fig.~\ref{figure:mapping}(a). 
The inset of Fig.~\ref{figure:mapping}(a) shows
$ \ln \langle t_b \rangle$ vs. $A$, from which 
a change in the slope (from $1.05$ to $2.61$) can be easily identified
when $A$ increases beyond about $2.6~k_{\rm B}T$. 
A corresponding exponential fit to $\langle t_b \rangle \sim A$
is also shown in the main panel and
indicates that the bond life time $\langle t_b \rangle$ starts to grow
faster with $A$ at $A/k_{\rm B}T \gtrsim 2.6$.

Figure~\ref{figure:BondEnergy} shows that the mean energy $U_B$ per bond
is less than the full bonding value $4A$ because of thermal fluctuations.
More data on $U_B$ at various $A$'s are shown in Fig.~\ref{figure:mapping}(b).
In addition to calculating $U_B$ from dimer simulations, 
we also did calculations for a starting state of a ring of 13 monomers
bonding laterally as a check for many-body effects and found none. 
The energy per bond did not depend on the starting state 
as seen in Fig.~\ref{figure:mapping}(b).
Data in Fig.~\ref{figure:mapping}(b) indicates
a best fit $U_B = 4*A-3.41~k_{\rm B}T$.
The slope has the expected value $4$ that is independent of temperature $T$.
The negative intercept $-3.41~k_{\rm B}T$ reflects the effect of 
thermal motions.

\begin{figure}[ht]
\centering
\includegraphics[width=2.75in]{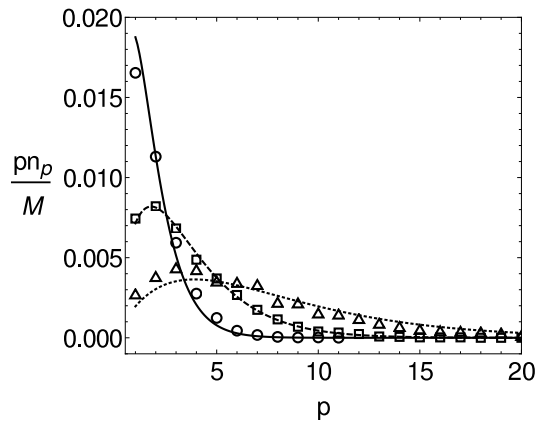}
\caption{The distribution of $p$-segment chains for $A_L=0$
and various $A_V$'s: 
$A_V=4.4~k_{\rm B}T$ (circles); 
$A_V=4.8~k_{\rm B}T$ (squares); 
$A_V=5.2~k_{\rm B}T$ (triangles). 
Lines are calculated from Eq.~(\ref{EqChainDist}) and
Eq.~(\ref{EqAssemCond}) with $\Delta g = -(4*A_V-13.03~k_{\rm B}T)$.
}
\label{figure:chaindist}
\end{figure}

To establish the mapping between $A$ and $\Delta g$,
we calculate the
chain distributions $n_p$ of $p$-segment chains for systems along
the line $A_L=0$, i.e., the straight polymerization case.
Along this line, dimers and short oligomers start to appear
when $A_V\gtrsim 3~k_{\rm B}T$, but
longer chains will not form significantly 
until $A_V\gtrsim 4\sim 5~k_{\rm B}T$.
The distributions of $n_p$ are calculated in MD simulations at
various $A_V$'s and compared with predictions of the thermodynamic theory 
(Eq.(\ref{EqChainDist}));
in the latter the dependence on $\Delta g$ enters in.
The results for $pn_p$ vs. $p$ 
at $A_V = 4.4$, $4.8$, and $5.2~k_{\rm B}T$ 
are shown in Fig.~\ref{figure:chaindist}. 
Chain distributions shown here are for systems starting with all monomers free.
To verify that they actually represent the equilibrium distribution 
at the corresponding $A_V$, we did another type of simulations where
the starting state was many pre-formed tubes.
For the values of $A_V$ used here, these tubes started to disassemble into
filaments, oligomers, and monomers with time. 
It was confirmed that these systems reached the same final distribution of $n_p$
as those reported here starting with free monomers.

The theoretical lines from Eq.~(\ref{EqChainDist}) with various $\Delta g$'s 
are also shown in Fig.~\ref{figure:chaindist} to compare with the MD results.
We found that the agreement is quite satisfactory 
if we use the the relation
\begin{equation}
\begin{array}{lll}
-\Delta g &=& U_B - 9.62~k_{\rm B}T \\
&=& 4*A-13.03~k_{\rm B}T,
\end{array}
\label{eq:dg}
\end{equation}
which effectively introduces
a minimal energy of $9.62~k_{\rm B}T$ 
for the bonding interactions of two monomers to form a dimer.
With this mapping, the thermodynamic theory 
accurately describes the straight polymerization process.
It is also found that assembly along the line $A_V=0$ 
is well described by the thermodynamic theory.
Further note that as $A_V$ is increased, the maximum of $n_p$ shifts
to larger $p$, which validates the criterion (Eq.~(\ref{phasecriterion}))
we employed earlier to determine the critical strength, $\Delta g_c(p)$, 
for the transition to $p$-segment chains.

The physical origin of the minimal energy $\sim 13~k_{\rm B}T$
for stable bonding incorporates three factors.
For any bond to occur, the bonding energy must be stronger than the
entropic cost which is $3~k_{\rm B}T$.
Even for bonding energies above this, there is a time in which the
dimer will dissociate (see Fig. \ref{figure:mapping}).
For growth of a chain, let alone a tubule, the stability time must be
greater than at least the collision time, which is the minimal time required 
for an additional monomer to bind.
The collision time depends on the concentration.
Thus, the value of $A$ determined here is only valid for the single 
concentration treated here.
Similar results have been found in the calculation 
for a simpler patchy particle model by Sciortino {\it et al.}.\cite{SciortinoJCP07}
Their work treats a hard core sphere with two short-ranged attractive spots 
at the particle poles, which can be treated analytically.
Figure 4 in Ref.~36 shows the polymerization transition 
as a function of the well depth and density.
For this simple model, the calculations can be done for a wide range of densities.
The value of the well depth necessary for polymerization depends logarithmically on the density.
At a density of 0.01, the minimum well depth for polymerization is about $10~k_{\rm B}T$.
In our simulations, along $A_L=0$ polymerization occurs
at about $A=3.3~k_{\rm B}T$ or a (surface-surface) well depth of 13.2 $k_{\rm B}T$.
The two systems are geometrically different in multiple ways and should not
produce identical values.
However, both produce values that are many $k_{\rm B}T$ beyond 3 $k_{\rm B}T$, which
is related to the time scales of diffusion and dissociation.
The translational diffusion times should be very similar for the two systems.
Because our wedge monomers have an orientation dependence, the rotational
diffusion times for binding will be longer than the sticky sphere model, which
will result in a larger value for the well depth as we find.
Our flat surface also causes more loss of entropy upon binding than for two spheres, 
which after binding still have rotational degrees of freedom. 
The binding energy for the wedge system will have to 
have an additional energy to compensate for this relative loss of the entropy.
Overall, the large well depth found in our wedge system is 
consistent with expectations and the magnitude is reasonable for such a system.
Furthermore, Erickson also estimated the work required to immobilize a subunit in a dimer
or polymer and obtained a range from 11.7 to 18.4 $k_{\rm B}T$,\cite{EricksonJMB89}
within which the minimal energy calculated here is located.

With the relationship between $A$ and $\Delta g$, 
we have calculated boundaries between the different structures found
in Fig. \ref{figure:phase} using the thermodynamic theory with 
some simple but ${\it ad~hoc}$ arguments.\cite{BruinsmaPRL03}
In the thermodynamic theory we obtained the critical binding energy
$\Delta g_c (p)$ that determines the assembly transition 
to $p$-segment chains. 
From $\Delta g_c (p)$, we
can easily compute the corresponding soft-potential strength $A_c$ 
at the transition through $\Delta g_c(p)=-(4*A_c-13.03~k_{\rm B}T)$ 
(for simplicity the energy unit $k_{\rm B}T$
is not always explicitly included in the following discussion).

When $A_L$ is small, no lateral bonding occurs and the problem
reduces to a living polymerization phenomenon. 
As discussed in Sec.~\ref{sec:theory}, the onset of assembly
is identified as when half of the monomers end up 
in the assembled state. This leads to a critical 
bonding energy $\Delta g_{cb} = -4.5~k_{\rm B}T$, 
corresponding a soft potential strength $A_{cb}=4.4~k_{\rm B}T$.
For $A_V < 4.4$ k$_B$T, most of the system is in the single monomer state.
Once $A_V > 4.4$ k$_B$T, significant self-assembly occurs with most of the
monomers now in some $n$-mer.
Similarly, when $A_V$ is small there is no vertical bonding and 
significant self-assembly of monomers into partial rings through lateral 
bonding occurs when $A_L > 4.4~k_{\rm B}T$. 
In Fig. \ref{figure:phase} the lines at $A_L = 4.4~k_{\rm B}T$ 
and at $A_V = 4.4~k_{\rm B}T$ are from this calculation and
they correspond well with
the boundaries between the oligomer state and A state, and between
the oligomer state and the F state as determined from the MD simulations.

In the region corresponding to the single monomer M state
(including the oligomer states $O_L$ and $O_V$), 
there is a sub-region bounded by the dashed-line parts 
of the two curves that separate the possible tube state
from other states. Inside this sub-region, monomers can only form 
unstable dimers from time to time and the number
of such dimers is small at any instant.
The boundaries of this region roughly corresponds to
$\Delta g<0$, which from Eq. (\ref{eq:dg}) is  $A < 3.3~k_{\rm B}T$.
These boundaries are also consistent
with the earlier estimate of $A$ for a stable bond
from the calculation of bond life time $t_b$ as shown in Fig.~\ref{figure:mapping}(a), 
which shows $\tau_b$ becomes nonzero near $3~k_{\rm B}T$.
A better estimate will be calculated below from the assembly kinetics
and it also indicates that 3.3 \kt{} is the boundary value.

Full ring formation determines the boundary between the A and R states.
Since an ideal ring constrains $p$ to be 13, 
we can use the condition $n_p(p=13) = 1$ to estimate the lower bound of this boundary.
This gives us a critical soft potential strength $A_L=5~k_{\rm B}T$
which is shown as a line in Fig. \ref{figure:phase} and fits the
simulations data for boundary with the R state.

At $\Delta g = \Delta g_c (p)$, the $p$-segment chains (or partial rings)
dominate. Then in order for these chains to assemble into tubes,
the bonding potential that holds two chains side by side
has to be larger than $9.62/p$ since there are $p$ side-to-side bonds, 
and this leads to $A > (3.41+9.62/p)/4$, where everything is in the unit of \kt.
The curves given by $\Delta g_c (p)=-(4*A_1-13.03)$
and $A_2 > (3.41+9.62/p)/4$ with
$(A_1,A_2)=(A_L,A_V)$ or $(A_1,A_2)=(A_V,A_L)$
determines the boundary between the F and S states 
and between the A/R states and the T state, 
respectively. 
These are the curved lines in Fig.~\ref{figure:phase}.

As shown in Fig.~\ref{figure:phase}, there is a remarkable agreement 
between the structure diagram calculated from MD simulations
and the prediction of the thermodynamic theory.
The predicated boundaries at M, F, S, A, R, and T states
all fit the simulation results well.
This indicates that the main feature of 
the assembly behavior of wedge-shaped monomers
is captured by the simple free energy developed in Sec.~\ref{sec:theory}.
Figure~\ref{figure:phase} further shows
that the kinetically trapped states (the S, C, K states) 
dominate when $A_V$ is large.
The thermodynamic theory can not capture this nonequilibrium behavior.

\subsection{Assembly Kinetics}

\begin{figure}[ht]
\centering
\includegraphics[width=2.5in]{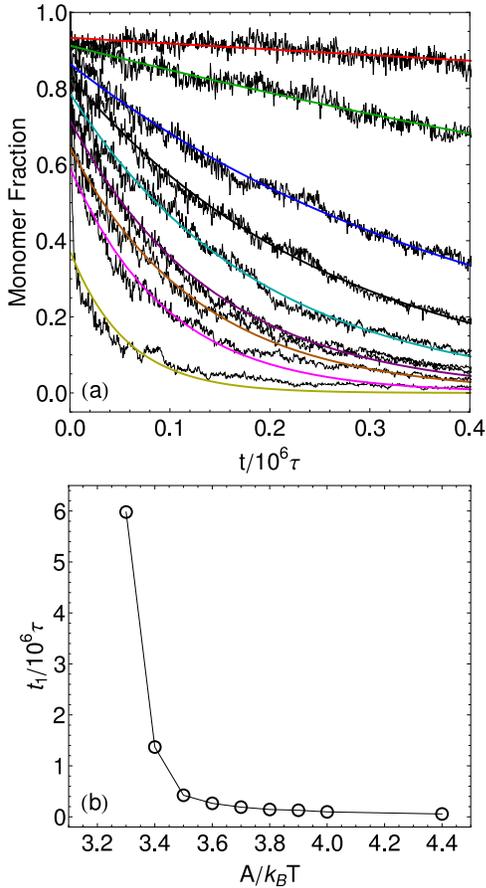}
\caption{(Color Online) (a) The fraction of free monomers vs. time for $A_L=A_V=A$. 
For the first eight curves from top to bottom, 
$A$ increases from $3.3~k_{\rm B}T$ to $4.0~k_{\rm B}T$ in increments of $0.1~k_{\rm B}T$.
The bottommost curve is for $A=4.4\epsilon$. The smooth lines are 
the fits to $a* \exp ( -t/t_1 )$. 
(b) The fitting parameter $t_1$ vs. $A$. The line is a guide to the eye.}
\label{figure:kinetics}
\end{figure}

We have shown that wedge-shaped monomers can self-assemble into various structures,
depending on the strengths of the lateral and vertical bonding interaction. 
When we start with a system with all monomers unbonded, the number of free monomers
decreases over time with the progress of self-assembly. 
The rate of consumption of free monomers
is controlled by the strength of the bonding potential.
In Fig.~\ref{figure:kinetics}(a),
results for the number of free monomers $n_1$ 
normalized by the total number of monomers ($500$ here) is plotted against time $t$ for
$A_L=A_V=A$, where $A$ ranges from $3.3$ to $4.4~k_{\rm B}T$.
The strength $3.3~k_{\rm B}T$ is close to the threshold for
assembly, and the corresponding $n_1$ decreases very slowly on this time scale,
though at a later time it will eventually drops below 50\% of starting free monomers.
At $A=3.4~k_{\rm B}T$, $n_1$ clearly decreases with time but many free monomers remain at the
end of the time range shown here.
As $A$ increases, the rate of reduction becomes larger.
At $A=4.4~k_{\rm B}T$, $n_1$ decreases to $0$ over a period of only $3\times 10^5\tau$,
which is an order of magnitude shorter than the typical time
of MD runs used to calculate the structure diagram.

Beyond an initial drop in $n_1$ at very short times,
the $n_1$ vs. $t$ curves can be fit to a functional form $a*\exp (-t/t_1)$, 
with $a$ and $t_1$ as the fitting parameters. 
The time decay constant for the single monomer count is $t_1$.
This time constant is for data beyond a very short initial time, 
when $n_1(t)$ drops sharply. 
From Fig.~\ref{figure:kinetics}(a) it is clear that even for the $A=3.3\,k_{\rm B}T$ case the
limit to $t=0$ is $n_1 < 1$.
The fits are shown as smooth lines in Fig.~\ref{figure:kinetics}(a).
The plot of the time constant $t_1$ vs. $A$ in Fig.~\ref{figure:kinetics}(b)
shows that $t_1$ appears to reach a critical value near $A \simeq 3.2~k_{\rm B}T$,
which is consistent with the earlier estimate of $3.3\,k_{\rm B}T$ as the threshold for 
some dimers to form.

\subsection{Helicity of Assembled Tubes}

\begin{figure}[ht]
  \subfigure[]{
    \begin{minipage}[b]{0.15\textwidth}
      \centering
      \includegraphics[width=1.1in]{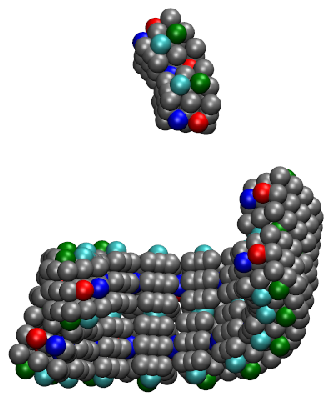}
     \end{minipage}}%
  \subfigure[]{
    \begin{minipage}[b]{0.18\textwidth}
      \centering
      \includegraphics[width=1.0in]{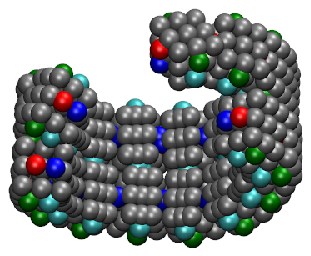}
     \end{minipage}}%
  \subfigure[]{
    \begin{minipage}[b]{0.14\textwidth}
      \centering
      \includegraphics[width=0.95in]{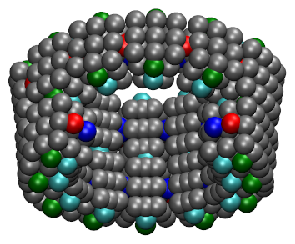}
     \end{minipage}}\\
  \subfigure[]{
    \begin{minipage}[b]{0.26\textwidth}
      \centering
      \includegraphics[width=1.0in]{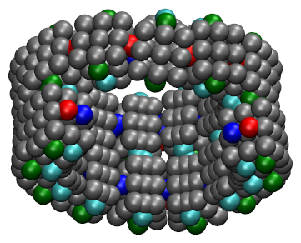}
     \end{minipage}}%
  \subfigure[]{
    \begin{minipage}[b]{0.2\textwidth}
      \centering
      \includegraphics[width=1.1in]{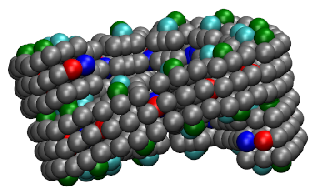}
     \end{minipage}}
\caption{Consecutive snapshots at $A_L=4.4~k_{\rm B}T$ 
and $A_V=2.6~k_{\rm B}T$ show the formation 
of a closed ring ((a)-(c)) by capturing a dimer
and the subsequent relaxation ((d)-(e)) into a helical structure.}
\label{figure:snapshotone}
\end{figure}

The visualization of the formation of an assembled structure
reveals important aspect of the assembly process.
The consecutive snapshots of assembly of one tubule in the system for 
Fig.~\ref{figure:tubules} shown in 
Fig.~\ref{figure:snapshotone} illustrate an interesting example.
In Fig.~\ref{figure:snapshotone}(a), two partial rings stacking vertically
have already formed, and a dimer is approaching this cluster.
In Fig.~\ref{figure:snapshotone}(b), the dimer has collided and bound
to the partial double ring.
In Fig.~\ref{figure:snapshotone}(c), the binding of the dimer results in
the formation of a full ring.
Then in Fig.~\ref{figure:snapshotone}(d), one wedge in the closed ring
oscillates noticeably and
eventually in Fig.~\ref{figure:snapshotone}(e), this wedge separates
from its neighbor and slips to bond with the wedges in the partial ring
below.
In this manner a helical ring is formed.
These helical structures typically have 12 or 13 monomers per turn.
This particular structure grows into a larger helical tubule with several
complete turns before the simulation ends.

\begin{figure}[htb]
  \subfigure[]{
    \begin{minipage}[b]{0.15\textwidth}
      \centering
      \includegraphics[width=1.0in]{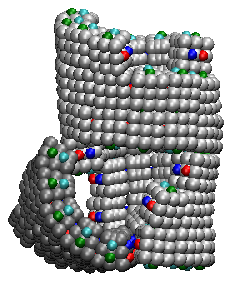}
     \end{minipage}}%
  \subfigure[]{
    \begin{minipage}[b]{0.15\textwidth}
      \centering
      \includegraphics[width=1.0in]{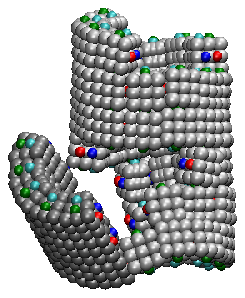}
     \end{minipage}}%
  \subfigure[]{
    \begin{minipage}[b]{0.15\textwidth}
      \centering
      \includegraphics[width=0.85in]{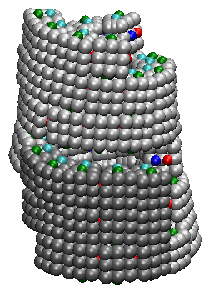}
     \end{minipage}}\\
  \subfigure[]{
    \begin{minipage}[b]{0.15\textwidth}
      \centering
      \includegraphics[width=0.7in]{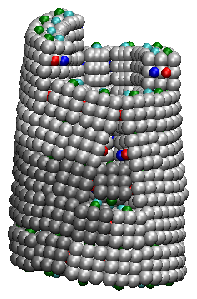}
     \end{minipage}}%
  \subfigure[]{
    \begin{minipage}[b]{0.15\textwidth}
      \centering
      \includegraphics[width=0.7in]{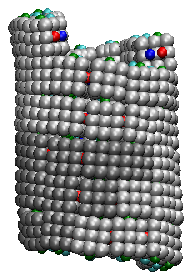}
     \end{minipage}}%
  \subfigure[]{
    \begin{minipage}[b]{0.15\textwidth}
      \centering
     \includegraphics[width=0.7in]{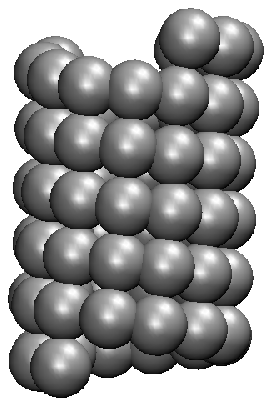}
     \end{minipage}}
\caption{Consecutive snapshots show the capture of a cluster by
a partial tube and the formation of a closed one ((a)-(f)). 
The tube subsequently relaxes into a helical structure ((g)-(i)).
In (j), each monomer is represented by a sphere
to illustrate clearly the helical feature of the tube.}
\label{figure:snapshottwo}
\end{figure}

Another interesting assembly dynamics involves the formation of tubes
not by the addition of monomer and dimers, but by collisions between
large clusters.
Figure~\ref{figure:snapshottwo} shows an example of this dynamics; the
two clusters are displayed through two different shades of gray.
In Fig.~\ref{figure:snapshottwo}(a), the larger cluster contains two
stacked rings (nonhelical) in the top half and 3 stacked arcs
on the bottom half.
This cluster is colliding with a curved sheet in this first image.
In Fig.~\ref{figure:snapshottwo}(b) the curved sheet rotates and orients
to fit into the missing bottom part of the larger cluster.
In Fig.~\ref{figure:snapshottwo}(c) the two clusters begin to merge. 
Note that the right hand side of the curved sheet (dark cluster) is straight
as this point.
In Fig.~\ref{figure:snapshottwo}(d), the two clusters have merged and undergone further 
rearrangement.
In particular, the wedge monomers on the right hand side, which in Fig.~\ref{figure:snapshottwo}(c)
are straight, on top of each other, have now translated in a staircase fashion.
Figure \ref{figure:snapshottwo}(e) shows that the triangular gap above one of the dark wedges has
healed and the whole combined structure is now a helical tube.
The final image in Fig.~\ref{figure:snapshottwo}(f) shows the wedges as spheres, clearly displaying a
well formed helical tube.

The geometry of our wedges are designed to form a 13-monomer ring
and then rings can stack to form tubes.
Though we do observe nonhelical tubes with 13 filaments, we also
see nonhelical tubes with only 12 monomers in each of their rings.
However, helical tubes containing 12 or 13 filaments
are more frequently found in our simulations.
We have directly compared the total potential energy of two preformed tubes of 
the same number of monomers with one helical and the other not.
The nonhelical tube does have the lower energy.
However, the difference in energy is small and about equal to the 
fluctuation in energy.
Thus, helical structure is sufficiently close in energy that it is not
surprising that we see so many.

\section{Discussion}

With dynamics that involves assembly and disassembly of structures, achieving equilibrium
in a simulation is typically a difficult task.
This is true in some parts of the structure diagram, but the boundaries between the structures
have been well determined at least to the precision that we desire.
The decay in the number of single monomers (Fig.~\ref{figure:kinetics}) shows that
the self-assembly process is very slow near the onset of assembly.
We can not run the simulations at $A=3.3$ \kt{} for long enough times to achieve 
equilibrium distributions.
However, by examining the kinetics as in Fig. \ref{figure:kinetics}(b), we know that
there is a boundary near 3.3 \kt{}.
The data presented here provides sufficient information for anyone interested in more
details of a particular structure to know what parameter range to focus on, which is
our interest.
We have achieved equilibrium distributions for some parts of the structure diagram.
In particular, the distributions in Fig. \ref{figure:chaindist} are equilibrium values.
We demonstrated equilibration by performing simulations of assembly (starting from monomers)
and of disassembly (starting from tubules).

At large values of $A$, kinetics tends to be the driver.
For the kinetically trapped region, the main point is that this is a region
to avoid when trying to form tubules.
Even experimentally, where much longer time scales can be easily reached, there
is no advantage to work in this region, because the kinetically trapped structures become trapped at relatively 
short times.

Experimentally, the advantage of longer equilibration times is used coincidentally with
lower concentrations.
In this work, we have kept the concentration constant in order to keep the number of 
variables manageable.
The choice of the particular concentration was a compromise between lower concentrations which 
would yield better assembly and higher concentrations at which assembly progresses faster.
It is apparent that the structure diagram will shift when concentration is varied. 
However, the thermodynamic theory predicts that the binding energy $\Delta g$ only
depends on concentration logarithmically.\cite{SciortinoJCP07} 
Thus the qualitative picture of various structures presented 
here should still remain valid, with some minor quantitative changes in the 
bond strengths required for tubule assembly.

Our results indicate that tubes are more easily and better formed with few defects
when the lateral binding
is slightly stronger than the vertical binding, i.e., $A_L \gtrsim A_V$. 
The desired values of $A_L$ is around 4 $k_{\rm B}T$ and $A_V$ 
around 3 $k_{\rm B}T$, corresponding to bonding energies about
13 $k_{\rm B}T$ laterally and 9 $k_{\rm B}T$ vertically, respectively.
These numbers should serve as reasonable guides for experimental design
of wedge-shaped molecules that can self-assemble into tubular structures.
We have also found that very long tubes can form 
when $A_V = 3.9$ $k_{\rm B}T$ and $A_L = 2.6$ $k_{\rm B}T$. 
In this case, only a few tubes and many monomers coexist 
almost without intermediate clusters. 
Tubes containing over 300 monomers and having greater than 20 turns have been assembled
in the simulations. However, tubes grow very slowly in this case.

There has been significant interest in modeling MT assembly kinetics and 
dynamics\cite{ChenPNAS85,BayleyJCS90,MartinBJ93,FlyvbjergPRE96,VanBurenPNAS02,WuPLOS09,PiettePLOS09}, 
including treating mechanical aspects\cite{MolodtsovBJ05,VanBurenBJ05}.
In Ref.~51, VanBuren {\it et al.} estimated 
bond energies within the MT lattice.
They predicted that the standard free energy gain 
of the formation of a vertical bond is 6.8 to 9.4 $k_{\rm B}T$
including the entropy contribution of immobilizing a dimer in the MT lattice.
Using our estimate of entropy loss of a bond formation as 9.6 $k_{\rm B}T$,
this corresponds to an energy of 16.4 to 19 $k_{\rm B}T$ for a vertical bond, 
which is translated to $A_V=5.0\sim 5.6$ $k_{\rm B}T$ in our model.
They also estimated that 
a lateral bond is much weaker with an energy of 3.2 to 5.7 $k_{\rm B}T$,
i.e., $A_L=1.7\sim 2.3$ $k_{\rm B}T$.
In our structure diagram (Fig.~\ref{figure:phase}), 
the region encircled by these values of $A_L$ and $A_V$ accommodates large sheets.
Though the thermodynamic theory predicts that the lowest energy state
is for these sheets to close into tubes, such closure is hindered
by the quick formation of many sheets and depletion of free monomers,
and difficulty of two or more sheets to join together to fuse into tubes
because of structural mismatch.
Only on a time scale much longer than the one achievable with MD,
can bond breaking and structural adjustment occur to allow tubes to form.
Also note that VanBuren {\it et al.} assumed that all the free energy 
of immobilizing a dimer is provided by the vertical bonding.
If instead we assumed that this free energy cost is jointly compensated by both
the vertical and the lateral bonds, then the lateral bond energy (i.e., $A_L$) would 
become larger while the vertical one (i.e., $A_V$) smaller, 
which moves the estimate of bond energies of a MT lattice closer
to the region where we observe tube formation in our simulations.

In a recent stochastic simulation work of Wu {\it et al.} that studied the intermediate sheet structure
found during MT assembly, two types of lateral bonds and one type of vertical bonds
were assumed to have energies of 13 to 17.5 $k_{\rm B}T$ and 19 $k_{\rm B}T$, respectively.\cite{WuPLOS09}
These correspond to $A_L=4.1\sim 5.2$ $k_{\rm B}T$ and $A_V=5.6$ $k_{\rm B}T$.
The lateral bond strength is inside the range found in this paper where tubular structures are formed,
but the vertical bond strength is in the region 
where kinetically trapped structures are found in our simulations. 
However, the monomer concentration in their work is an order of magnitude lower than 
ours, which implies that a larger value of $A_V$ is required for tube formation
in their simulations because of the increase of diffusion time and the decrease of collision rate.
Furthermore, in stochastic simulations it is much easier for a system to escape from
kinetic traps and reach a thermodynamic equilibrium state where tubes are expected to form. 
We thus conclude that the parameter set used by Wu {\it et al.} is 
consistent with the range of bond strengths suitable for tube formation found in this paper.

Besides the interaction strengths, the geometry of the monomer is also important
in determining the self-assembled structure.
Surprisingly, we have found that helical tubes are more common than nonhelical tubes.
The nonhelical tube has the lowest energy, but the difference between the two states
is small, which is found less than the fluctuation in the energy.
Thus helical structures emerge with entropy as a driven force.
The formation of the helical tubes occurs in a variety of ways as shown in Figs. 
\ref{figure:snapshotone} and \ref{figure:snapshottwo}.
Once the helical structure has formed for a length of about one turn, then it is
stable as helical and grows from the ends.
The collisional reformation seen in Fig. \ref{figure:snapshottwo} is not a common mechanism.
The most common mechanism appears to be the formation of a curved sheet of a few layers 
that has sufficient filaments to bend and touch itself.
In most cases the self-contact is with a twist yielding helical tubes.
This mechanisms has been seen in other systems such as lipid nanotubes.

We mentioned that when $A_L > 5\,k_{\rm B}T$ and $A_V < 1\,k_{\rm B}T$ 
rings with various number ($11 \sim 16$) of monomers can form. 
In our model, 13 monomers fit to make a perfect ring. So rings
containing more than 13 monomers have to take a somewhat out-of-plane shape 
because our monomer is a rigid composite body.
This shape makes tube formation difficult, 
though tubes do form when $A_V$ is increased above $1\,k_{\rm B}T$ in this range of $A_L$.
However, only tubes with 12 or 13 filaments, either helical or nonhelical, are observed;
the majority has 13 filaments.
This is slightly different from the case of MT assembly {\it in vitro},
where the protofilament number of MTs varies between 10 and 15, with the vast
majority having 14 protofilaments.\cite{Desai97}
The smaller range found in our simulations
is clearly due to the fact that our monomer is rigid.
The distortion of the rings containing other than 12 or 13 monomers 
is incompatible with the geometry of a tube, and the monomer cannot adjust its shape to
reduce the distortion of the ring.

We have used the wedge shape to promote formation of the tubule structure.
We believe that this is not necessary.
A more spherical or ellipsoidal shape could be used with the lateral interaction sites
being located off center in order to promote ring formation.
However, changing to a more curved shape would have some consequences which would require
adjustments.
On the one hand, the flat wedge surfaces yield a tighter fit between bound monomers, which raises the entropic
cost and must be compensated for in the binding energy.
On the other hand, a more flexible bond will be more unstable 
(lower barrier) and potentially allow other composite geometries.
We have seen in the helical twist that even small variations in the geometry 
between two bound monomers can yield very different structures on larger scales.
The wedge shape brings some simplification which enables such difficult simulations, 
but some of the future work will need to match more closely the actual molecular building blocks.

The goal of this work is to study a model monomer designed to form tubules, 
which is motivated by an interest in developing artificial MTs.
This model can be further developed to study real MTs and 
it is interesting to see how well the model compares to MT assembly 
and structure, though only a few aspects of tubulin dimers are incorporated in the model, 
which will limit a direct comparison.
The basic result of forming tubules and finding the appropriate range of interaction strengths, 
which was discussed above, is the necessary first step.
The fact that helical tubules are formed in the simulations is a pleasant surprise, 
although one that is understandable.
Furthermore, it indicates that some features in the monomer 
must be adjusted in order to control the helical assembly.
The preference for the lateral strength to be greater than 
the vertical strength is interesting and deserves more study.
The results show the importance of nucleation in MT assembly, 
but generally speaking this is not a surprise in biological systems 
which tend to be controlled in a systematic manner.

\section{Conclusions}

We have developed a minimal model of tube formation from wedge-shaped monomers.
Our simulations have identified a rich structure diagram of the assembly,
which agrees well with the prediction of a simple Flory-Huggins lattice theory.
Results indicate that tubes form when both the attractive
lateral and vertical binding
interaction energies are at the scale of $\sim 10\,k_{\rm B}T$.
Tubes with fewer defects are more easily formed when the lateral interaction
is slightly stronger than the vertical one.
The stacking of rings is less susceptible to kinetic traps than the alignment and piecing
together of filaments.
There is more than one assembly mechanism for tube formation.
We have observed that in addition to the main growth mechanism by absorbing free monomers,
tubes can also grow through capturing dimers and oligomers, or even through
collisions between clusters.
Our results also reveal that helical tubes
are more frequently formed than nonhelical tubes, despite the fact that our monomer geometry
and interaction potentials favor nonhelical structures, which have the lowest energy.
Part of the reason for this is that the energy difference is small, which allows transitions
between rings and helices of a single turn, for example.

This work sets the stage for further model developments 
that will incorporate additional features necessary for simulating 
other aspects of MTs such as strong depolymerization.
Our interest is more in developing design rules for constructing artificial MTs, but that
involves understanding MTs in a more general manner.
The present results indicate the best range of interaction strengths to achieve 
artificial tubule formation.
In addition, the simulations show that cells have to control multiple aspects of
the formation dynamics to prevent defects from forming in the assembly of MTs.
Tubules can collide and bind, which tends to form structures with defects.
In the helical tubules, we find some with a pitch of 2 monomers instead of 1.
All these factors suggest that instability inherent within MTs 
is also a mechanism to handle unwanted, poorly growing structures.

\section*{Appendix}

In this appendix, we give a detailed derivation of the free energy used in the main text. 
Since chains of different lengths are distinguishable,
we can first place the 1-segment chains (leftover monomers) onto the lattice, 
then place the 2-segment chains (dimers), and so on. Let $\xi_{p,i}$ 
be the number of ways to place the $i$-th $p$-segment chain, and
$W_p$ be the total number of ways to place all $p$-segment chains,
and taking into account the indistinguishability amongst the $p$-segment chains,
we obtain
\begin{equation}
W_p = \frac{1}{n_p !}\prod_{i=1}^{n_p}\xi_{p,i}.
\label{EqNumPSeg}
\end{equation}
The number of occupied sites by all chains up to but not including the $p$-segments is
\begin{equation}
M_p = \sum_{j=1}^{p-1}jn_j,~~~{\rm for~} p \ge 2
\end{equation}
Note that $M_1=0$. 
There are $M-M_p$ cells available for the first $p$-segment chain.
When we start placing the $i$-th $p$-segment chain, there are
$M-M_p-(i-1)p$ cells empty for its first segment (monomer). 
Since the wedge monomer is an anisotropic object, 
its orientation contributes to the entropy of the system.
In the lattice model, the monomer has $z$ possible orientations, 
where $z$ is the coordination number of the lattice.
For modeling the wedge monomers, $z=6$ for a cubic lattice is most appropriate.
Thus the number of different ways to place the first segment
of the $i$-th $p$-segment chain is 
\begin{equation}
K_1 = z\left[ M-M_p-(i-1)p \right].
\end{equation}
Since each monomer can form two bonds on its opposite faces in linear chains,
the second segment has to be placed in $2$ neighboring cells 
of the first segment where bonding is possible.
Using mean-field approximation, 
the probability for one such cell to be empty
is $\left[ M-M_p-(i-1)p-1 \right]/M$.
The number of ways to place the second segment of the
$i$-th $p$-segment chain is then 
\begin{equation}
K_2 = \frac{2\left[ M-M_p-(i-1)p-1 \right]}{M}.
\end{equation}
Now the third segment can only be placed in 
the cell adjacent to the one that the second segment resides,
but opposite to the cell that the first segment resides.
The probability for this cell to be available is $\left[ M-M_p-(i-1)p-2 \right]/M$.
The number of ways to place the third or any segment $j\ge 3$ of the
$i$-th $p$-segment chain is
\begin{equation}
K_j = \frac{M-M_p-(i-1)p-(j-1) }{M}.
\end{equation}

From Eq.~(\ref{EqNumPSeg}) we obtain
\begin{equation}
\begin{split}
W_p & = \frac{1}{2}\frac{1}{n_p !}\prod_{i=1}^{n_p}\prod_{j=1}^{p} K_j \\
    & = \frac{1}{n_p !}\frac{z^{n_p}}{M^{n_p(p-1)}}
\frac{(M-M_p)!}{(M-M_{p+1})!}.
\end{split}
\end{equation}
Here the identity $M_{p+1}=M_p+p n_p$ is used.
The factor $1/2$ is included because there are $2$
ways to choose the first segment of a rigid linear chain,
and thus there is a $2$-fold degeneracy in the above counting
of ways to place chains.
Finally, the total number of different ways to place all chains is
\begin{equation}
\begin{split}
Z &= \prod_{p=1}^{p_{\rm max}} W_p \\
 &= \frac{M!}{ n_s ! \prod_{p=1}^{p_{\rm max}} n_p !}
\frac{z^{\sum_{p=1}^{p_{\rm max}}n_p}}{M^{\sum_{p=1}^{p_{\rm max}}n_p(p-1)}},
\end{split}
\end{equation}
where $n_s = M-\sum_{p=1}^{p_{\rm max}}pn_p=M-N$ is 
the number of solvent (empty) cells after all chains are placed.

The configurational entropy $S_c$ of the assembled system is
\begin{equation}
\begin{split}
S_c &= k_{\rm B} \ln  Z \\
&= k_{\rm B}\sum_{p=1}^{p_{\rm max}}\left( n_p \ln z - n_p  \ln \frac{n_p}{M} + n_p\right)
\end{split}
\end{equation}
where $k_{\rm B}$ is the Boltzmann constant,
and terms that only depend on $n_s$ and $M$ are discarded
since they are constants.

The Helmholtz free energy can be written as
\begin{equation}
\begin{split}
F &= E-TS_c \\
&= \frac{1}{\beta}
\sum_{p=1}^{p_{\rm max}}n_p\left( \beta (p-1) \Delta g - \ln z +  \ln \frac{n_p}{M} - 1\right),
\end{split}
\end{equation}
where $(p-1)\Delta g$ is the total energy gain to form a $p$-segment chain,
which has $p-1$ bonds, and $\beta = 1/k_{\rm B}T$.

\section*{ACKNOWLEDGMENTS}
Sandia is a multiprogram laboratory operated by Sandia Corporation, 
a Lockheed Martin Company, for the United States Department of Energy 
under Contract No. DE-AC04-94AL85000.
This research was supported by the U.S. Department of Energy, 
Office of Basic Energy Sciences, Division of Materials Sciences 
and Engineering under Award KC0203010.


\begin{thebibliography}{55}
\expandafter\ifx\csname natexlab\endcsname\relax\def\natexlab#1{#1}\fi
\expandafter\ifx\csname bibnamefont\endcsname\relax
  \def\bibnamefont#1{#1}\fi
\expandafter\ifx\csname bibfnamefont\endcsname\relax
  \def\bibfnamefont#1{#1}\fi
\expandafter\ifx\csname citenamefont\endcsname\relax
  \def\citenamefont#1{#1}\fi
\expandafter\ifx\csname url\endcsname\relax
  \def\url#1{\texttt{#1}}\fi
\expandafter\ifx\csname urlprefix\endcsname\relax\def\urlprefix{URL }\fi
\providecommand{\bibinfo}[2]{#2}
\providecommand{\eprint}[2][]{\url{#2}}

\bibitem[{\citenamefont{Mann}(2009)}]{MannNM09}
\bibinfo{author}{\bibfnamefont{S.}~\bibnamefont{Mann}},
  \bibinfo{journal}{Nature Mater.} \textbf{\bibinfo{volume}{8}},
  \bibinfo{pages}{781} (\bibinfo{year}{2009}).

\bibitem[{\citenamefont{{Special issue on ``Supramolecular chemistry and
  self-assembly"}}(2002)}]{science295}
\bibinfo{author}{\bibnamefont{{Special issue on ``Supramolecular chemistry and
  self-assembly"}}}, \bibinfo{journal}{Science} \textbf{\bibinfo{volume}{295}},
  \bibinfo{pages}{2395} (\bibinfo{year}{2002}).

\bibitem[{\citenamefont{Desai and Mitchison}(1997)}]{Desai97}
\bibinfo{author}{\bibfnamefont{A.}~\bibnamefont{Desai}} \bibnamefont{and}
  \bibinfo{author}{\bibfnamefont{T.}~\bibnamefont{Mitchison}},
  \bibinfo{journal}{Annu. Rev. Cell Dev. Biol.} \textbf{\bibinfo{volume}{13}},
  \bibinfo{pages}{83} (\bibinfo{year}{1997}).

\bibitem[{\citenamefont{Nogales}(2000)}]{NogalesAR00}
\bibinfo{author}{\bibfnamefont{E.}~\bibnamefont{Nogales}},
  \bibinfo{journal}{Annu. Rev. Biochem.} \textbf{\bibinfo{volume}{69}},
  \bibinfo{pages}{277} (\bibinfo{year}{2000}).

\bibitem[{\citenamefont{Polaske et~al.}(2010)\citenamefont{Polaske, McGrath,
  and McElhanon}}]{McElhanonMM10}
\bibinfo{author}{\bibfnamefont{N.}~\bibnamefont{Polaske}},
  \bibinfo{author}{\bibfnamefont{D.}~\bibnamefont{McGrath}}, \bibnamefont{and}
  \bibinfo{author}{\bibfnamefont{J.}~\bibnamefont{McElhanon}},
  \bibinfo{journal}{Macromolecules} \textbf{\bibinfo{volume}{43}},
  \bibinfo{pages}{1270} (\bibinfo{year}{2010}).

\bibitem[{\citenamefont{Nogales et~al.}(1998)\citenamefont{Nogales, Wolf, and
  Downing}}]{NogalesNature98}
\bibinfo{author}{\bibfnamefont{E.}~\bibnamefont{Nogales}},
  \bibinfo{author}{\bibfnamefont{S.~G.} \bibnamefont{Wolf}}, \bibnamefont{and}
  \bibinfo{author}{\bibfnamefont{K.~H.} \bibnamefont{Downing}},
  \bibinfo{journal}{Nature} \textbf{\bibinfo{volume}{391}},
  \bibinfo{pages}{199} (\bibinfo{year}{1998}).

\bibitem[{\citenamefont{Kollman et~al.}(2011)\citenamefont{Kollman, Merdes,
  Mourey, and Agard}}]{KollmanNRMCB11}
\bibinfo{author}{\bibfnamefont{J.~M.} \bibnamefont{Kollman}},
  \bibinfo{author}{\bibfnamefont{A.}~\bibnamefont{Merdes}},
  \bibinfo{author}{\bibfnamefont{L.}~\bibnamefont{Mourey}}, \bibnamefont{and}
  \bibinfo{author}{\bibfnamefont{D.}~\bibnamefont{Agard}},
  \bibinfo{journal}{Nature Rev. Mol. Cell Biol.} \textbf{\bibinfo{volume}{12}},
  \bibinfo{pages}{709} (\bibinfo{year}{2011}).

\bibitem[{\citenamefont{Zheng et~al.}(1995)\citenamefont{Zheng, Wong, Alberts,
  and Mitchison}}]{ZhengNature95}
\bibinfo{author}{\bibfnamefont{Y.}~\bibnamefont{Zheng}},
  \bibinfo{author}{\bibfnamefont{M.~L.} \bibnamefont{Wong}},
  \bibinfo{author}{\bibfnamefont{B.}~\bibnamefont{Alberts}}, \bibnamefont{and}
  \bibinfo{author}{\bibfnamefont{T.}~\bibnamefont{Mitchison}},
  \bibinfo{journal}{Nature} \textbf{\bibinfo{volume}{378}},
  \bibinfo{pages}{578} (\bibinfo{year}{1995}).

\bibitem[{\citenamefont{Pum and Sleytr}(1995)}]{Pum95}
\bibinfo{author}{\bibfnamefont{D.}~\bibnamefont{Pum}} \bibnamefont{and}
  \bibinfo{author}{\bibfnamefont{U.~B.} \bibnamefont{Sleytr}},
  \bibinfo{journal}{Colloids and Surfaces A} \textbf{\bibinfo{volume}{102}},
  \bibinfo{pages}{99} (\bibinfo{year}{1995}).

\bibitem[{\citenamefont{Jarosch et~al.}(2001)\citenamefont{Jarosch, Egelseer,
  Huber, Moll, Mattanovich, Sleytr1, and S\'{a}ra}}]{Jarosch01}
\bibinfo{author}{\bibfnamefont{M.}~\bibnamefont{Jarosch}},
  \bibinfo{author}{\bibfnamefont{E.~M.} \bibnamefont{Egelseer}},
  \bibinfo{author}{\bibfnamefont{C.}~\bibnamefont{Huber}},
  \bibinfo{author}{\bibfnamefont{D.}~\bibnamefont{Moll}},
  \bibinfo{author}{\bibfnamefont{D.}~\bibnamefont{Mattanovich}},
  \bibinfo{author}{\bibfnamefont{U.~B.} \bibnamefont{Sleytr1}},
  \bibnamefont{and} \bibinfo{author}{\bibfnamefont{M.}~\bibnamefont{S\'{a}ra}},
  \bibinfo{journal}{Microbiology} \textbf{\bibinfo{volume}{147}},
  \bibinfo{pages}{1353} (\bibinfo{year}{2001}).

\bibitem[{\citenamefont{Mertig et~al.}(2001)\citenamefont{Mertig, Wahl,
  Lehmann, Simon, and Pompe}}]{Mertig01}
\bibinfo{author}{\bibfnamefont{M.}~\bibnamefont{Mertig}},
  \bibinfo{author}{\bibfnamefont{R.}~\bibnamefont{Wahl}},
  \bibinfo{author}{\bibfnamefont{M.}~\bibnamefont{Lehmann}},
  \bibinfo{author}{\bibfnamefont{P.}~\bibnamefont{Simon}}, \bibnamefont{and}
  \bibinfo{author}{\bibfnamefont{W.}~\bibnamefont{Pompe}},
  \bibinfo{journal}{Eur. Phys. J. D} \textbf{\bibinfo{volume}{16}},
  \bibinfo{pages}{317} (\bibinfo{year}{2001}).

\bibitem[{\citenamefont{Bobeth et~al.}(2011)\citenamefont{Bobeth, Blecha,
  Bl\"{u}her, Mertig, Korkmaz, Ostermann, R\"{o}del, and
  Pompe}}]{BobethLangmuir11}
\bibinfo{author}{\bibfnamefont{M.}~\bibnamefont{Bobeth}},
  \bibinfo{author}{\bibfnamefont{A.}~\bibnamefont{Blecha}},
  \bibinfo{author}{\bibfnamefont{A.}~\bibnamefont{Bl\"{u}her}},
  \bibinfo{author}{\bibfnamefont{M.}~\bibnamefont{Mertig}},
  \bibinfo{author}{\bibfnamefont{N.}~\bibnamefont{Korkmaz}},
  \bibinfo{author}{\bibfnamefont{K.}~\bibnamefont{Ostermann}},
  \bibinfo{author}{\bibfnamefont{G.}~\bibnamefont{R\"{o}del}},
  \bibnamefont{and} \bibinfo{author}{\bibfnamefont{W.}~\bibnamefont{Pompe}},
  \bibinfo{journal}{Langmuir} \textbf{\bibinfo{volume}{27}},
  \bibinfo{pages}{15102} (\bibinfo{year}{2011}).

\bibitem[{\citenamefont{Korkmaz
  et~al.}(2011{\natexlab{a}})\citenamefont{Korkmaz, Ostermann, and
  R\"{o}del}}]{KorkmazNT11a}
\bibinfo{author}{\bibfnamefont{N.}~\bibnamefont{Korkmaz}},
  \bibinfo{author}{\bibfnamefont{K.}~\bibnamefont{Ostermann}},
  \bibnamefont{and}
  \bibinfo{author}{\bibfnamefont{G.}~\bibnamefont{R\"{o}del}},
  \bibinfo{journal}{Nanotechnology} \textbf{\bibinfo{volume}{22}},
  \bibinfo{pages}{095601} (\bibinfo{year}{2011}{\natexlab{a}}).

\bibitem[{\citenamefont{Sleytr et~al.}(2007)\citenamefont{Sleytr, Egelseer,
  Ilk, Pum, and Schuster}}]{Sleytr07}
\bibinfo{author}{\bibfnamefont{U.~B.} \bibnamefont{Sleytr}},
  \bibinfo{author}{\bibfnamefont{E.~M.} \bibnamefont{Egelseer}},
  \bibinfo{author}{\bibfnamefont{N.}~\bibnamefont{Ilk}},
  \bibinfo{author}{\bibfnamefont{D.}~\bibnamefont{Pum}}, \bibnamefont{and}
  \bibinfo{author}{\bibfnamefont{B.}~\bibnamefont{Schuster}},
  \bibinfo{journal}{FEBS J.} \textbf{\bibinfo{volume}{274}},
  \bibinfo{pages}{323} (\bibinfo{year}{2007}).

\bibitem[{\citenamefont{Korkmaz
  et~al.}(2011{\natexlab{b}})\citenamefont{Korkmaz, B\"{o}rrnert, K\"{o}hler,
  Mendes, Bachmatiuk, R\"{u}mmeli, B\"{u}chner, Eng, and
  R\"{o}del}}]{KorkmazNT11b}
\bibinfo{author}{\bibfnamefont{N.}~\bibnamefont{Korkmaz}},
  \bibinfo{author}{\bibfnamefont{F.}~\bibnamefont{B\"{o}rrnert}},
  \bibinfo{author}{\bibfnamefont{D.}~\bibnamefont{K\"{o}hler}},
  \bibinfo{author}{\bibfnamefont{R.~G.} \bibnamefont{Mendes}},
  \bibinfo{author}{\bibfnamefont{A.}~\bibnamefont{Bachmatiuk}},
  \bibinfo{author}{\bibfnamefont{M.~H.} \bibnamefont{R\"{u}mmeli}},
  \bibinfo{author}{\bibfnamefont{B.}~\bibnamefont{B\"{u}chner}},
  \bibinfo{author}{\bibfnamefont{L.~M.} \bibnamefont{Eng}}, \bibnamefont{and}
  \bibinfo{author}{\bibfnamefont{G.}~\bibnamefont{R\"{o}del}},
  \bibinfo{journal}{Nanotechnology} \textbf{\bibinfo{volume}{22}},
  \bibinfo{pages}{375606} (\bibinfo{year}{2011}{\natexlab{b}}).

\bibitem[{\citenamefont{Schnur}(1993)}]{Schnur93}
\bibinfo{author}{\bibfnamefont{J.~M.} \bibnamefont{Schnur}},
  \bibinfo{journal}{Science} \textbf{\bibinfo{volume}{262}},
  \bibinfo{pages}{1669} (\bibinfo{year}{1993}).

\bibitem[{\citenamefont{Orr et~al.}(1999)\citenamefont{Orr, Barbour, and
  Atwood}}]{Orr99}
\bibinfo{author}{\bibfnamefont{G.~W.} \bibnamefont{Orr}},
  \bibinfo{author}{\bibfnamefont{L.~J.} \bibnamefont{Barbour}},
  \bibnamefont{and} \bibinfo{author}{\bibfnamefont{J.~L.}
  \bibnamefont{Atwood}}, \bibinfo{journal}{Science}
  \textbf{\bibinfo{volume}{285}}, \bibinfo{pages}{1049} (\bibinfo{year}{1999}).

\bibitem[{\citenamefont{Bong et~al.}(2001)\citenamefont{Bong, Clark, Granja,
  and Ghadiri}}]{Bong01}
\bibinfo{author}{\bibfnamefont{D.~T.} \bibnamefont{Bong}},
  \bibinfo{author}{\bibfnamefont{T.~D.} \bibnamefont{Clark}},
  \bibinfo{author}{\bibfnamefont{J.~R.} \bibnamefont{Granja}},
  \bibnamefont{and} \bibinfo{author}{\bibfnamefont{M.~R.}
  \bibnamefont{Ghadiri}}, \bibinfo{journal}{Angew. Chem. Int. Ed}
  \textbf{\bibinfo{volume}{40}}, \bibinfo{pages}{988} (\bibinfo{year}{2001}).

\bibitem[{\citenamefont{Yan et~al.}(2004)\citenamefont{Yan, Zhou, and
  Hou}}]{Yan04}
\bibinfo{author}{\bibfnamefont{D.}~\bibnamefont{Yan}},
  \bibinfo{author}{\bibfnamefont{Y.}~\bibnamefont{Zhou}}, \bibnamefont{and}
  \bibinfo{author}{\bibfnamefont{J.}~\bibnamefont{Hou}},
  \bibinfo{journal}{Science} \textbf{\bibinfo{volume}{303}},
  \bibinfo{pages}{65} (\bibinfo{year}{2004}).

\bibitem[{\citenamefont{Hill et~al.}(2004)\citenamefont{Hill, Jin, Kosaka,
  Fukushima, Ichihara, Shimomura, Ito, Hashizume, Ishii, and Aida}}]{Hill04}
\bibinfo{author}{\bibfnamefont{J.~P.} \bibnamefont{Hill}},
  \bibinfo{author}{\bibfnamefont{W.}~\bibnamefont{Jin}},
  \bibinfo{author}{\bibfnamefont{A.}~\bibnamefont{Kosaka}},
  \bibinfo{author}{\bibfnamefont{T.}~\bibnamefont{Fukushima}},
  \bibinfo{author}{\bibfnamefont{H.}~\bibnamefont{Ichihara}},
  \bibinfo{author}{\bibfnamefont{T.}~\bibnamefont{Shimomura}},
  \bibinfo{author}{\bibfnamefont{K.}~\bibnamefont{Ito}},
  \bibinfo{author}{\bibfnamefont{T.}~\bibnamefont{Hashizume}},
  \bibinfo{author}{\bibfnamefont{N.}~\bibnamefont{Ishii}}, \bibnamefont{and}
  \bibinfo{author}{\bibfnamefont{T.}~\bibnamefont{Aida}},
  \bibinfo{journal}{Science} \textbf{\bibinfo{volume}{304}},
  \bibinfo{pages}{1481} (\bibinfo{year}{2004}).

\bibitem[{\citenamefont{Shimizu}(2005)}]{Shimizu05}
\bibinfo{author}{\bibfnamefont{T.}~\bibnamefont{Shimizu}},
  \bibinfo{journal}{Chem. Rev.} \textbf{\bibinfo{volume}{105}},
  \bibinfo{pages}{1401} (\bibinfo{year}{2005}).

\bibitem[{\citenamefont{Shimizu}(2008)}]{Shimizu08}
\bibinfo{author}{\bibfnamefont{T.}~\bibnamefont{Shimizu}},
  \bibinfo{journal}{Bull. Chem. Soc. Jpn.} \textbf{\bibinfo{volume}{81}},
  \bibinfo{pages}{1554} (\bibinfo{year}{2008}).

\bibitem[{\citenamefont{Zhang et~al.}(2007)\citenamefont{Zhang, Chen, and
  W\"{u}rthner}}]{Zhang07}
\bibinfo{author}{\bibfnamefont{X.}~\bibnamefont{Zhang}},
  \bibinfo{author}{\bibfnamefont{Z.}~\bibnamefont{Chen}}, \bibnamefont{and}
  \bibinfo{author}{\bibfnamefont{F.}~\bibnamefont{W\"{u}rthner}},
  \bibinfo{journal}{J. Am. Chem. Soc.} \textbf{\bibinfo{volume}{129}},
  \bibinfo{pages}{4886} (\bibinfo{year}{2007}).

\bibitem[{\citenamefont{Kwak et~al.}(2010)\citenamefont{Kwak, Shin, Seok, Kim,
  Ahmad, Geckeler, Seeck, Seo, Satijag, and Kubik}}]{Kwak10}
\bibinfo{author}{\bibfnamefont{B.}~\bibnamefont{Kwak}},
  \bibinfo{author}{\bibfnamefont{K.}~\bibnamefont{Shin}},
  \bibinfo{author}{\bibfnamefont{S.}~\bibnamefont{Seok}},
  \bibinfo{author}{\bibfnamefont{D.}~\bibnamefont{Kim}},
  \bibinfo{author}{\bibfnamefont{F.}~\bibnamefont{Ahmad}},
  \bibinfo{author}{\bibfnamefont{K.~E.} \bibnamefont{Geckeler}},
  \bibinfo{author}{\bibfnamefont{O.~H.} \bibnamefont{Seeck}},
  \bibinfo{author}{\bibfnamefont{Y.}~\bibnamefont{Seo}},
  \bibinfo{author}{\bibfnamefont{S.~K.} \bibnamefont{Satijag}},
  \bibnamefont{and} \bibinfo{author}{\bibfnamefont{S.}~\bibnamefont{Kubik}},
  \bibinfo{journal}{Soft Matter} \textbf{\bibinfo{volume}{6}},
  \bibinfo{pages}{4701} (\bibinfo{year}{2010}).

\bibitem[{\citenamefont{Hamley}(2011)}]{Hamley11}
\bibinfo{author}{\bibfnamefont{I.~W.} \bibnamefont{Hamley}},
  \bibinfo{journal}{Soft Matter} \textbf{\bibinfo{volume}{7}},
  \bibinfo{pages}{4122} (\bibinfo{year}{2011}).

\bibitem[{\citenamefont{Coleman et~al.}(2011)\citenamefont{Coleman, Beierle1,
  M, Maci\'{a}, Caroli, Mika, van Dijken, Chen, Browne, and
  Feringa}}]{Coleman11}
\bibinfo{author}{\bibfnamefont{A.~C.} \bibnamefont{Coleman}},
  \bibinfo{author}{\bibfnamefont{J.~M.} \bibnamefont{Beierle1}},
  \bibinfo{author}{\bibfnamefont{C.~A.~S.} \bibnamefont{M}},
  \bibinfo{author}{\bibfnamefont{B.}~\bibnamefont{Maci\'{a}}},
  \bibinfo{author}{\bibfnamefont{G.}~\bibnamefont{Caroli}},
  \bibinfo{author}{\bibfnamefont{J.~T.} \bibnamefont{Mika}},
  \bibinfo{author}{\bibfnamefont{D.~J.} \bibnamefont{van Dijken}},
  \bibinfo{author}{\bibfnamefont{J.}~\bibnamefont{Chen}},
  \bibinfo{author}{\bibfnamefont{W.~R.} \bibnamefont{Browne}},
  \bibnamefont{and} \bibinfo{author}{\bibfnamefont{B.~L.}
  \bibnamefont{Feringa}}, \bibinfo{journal}{Nature Nanotech.}
  \textbf{\bibinfo{volume}{6}}, \bibinfo{pages}{547} (\bibinfo{year}{2011}).

\bibitem[{\citenamefont{Caspar and Klug}(1962)}]{Caspar62}
\bibinfo{author}{\bibfnamefont{D.~L.~D.} \bibnamefont{Caspar}}
  \bibnamefont{and} \bibinfo{author}{\bibfnamefont{A.}~\bibnamefont{Klug}},
  \bibinfo{journal}{Cold Spring Harbor Symp. Quant. Biol.}
  \textbf{\bibinfo{volume}{27}}, \bibinfo{pages}{1} (\bibinfo{year}{1962}).

\bibitem[{\citenamefont{Bruinsma et~al.}(2003)\citenamefont{Bruinsma, Gelbart,
  Reguera, Rudnick, and Zandi}}]{BruinsmaPRL03}
\bibinfo{author}{\bibfnamefont{R.~F.} \bibnamefont{Bruinsma}},
  \bibinfo{author}{\bibfnamefont{W.~M.} \bibnamefont{Gelbart}},
  \bibinfo{author}{\bibfnamefont{D.}~\bibnamefont{Reguera}},
  \bibinfo{author}{\bibfnamefont{J.}~\bibnamefont{Rudnick}}, \bibnamefont{and}
  \bibinfo{author}{\bibfnamefont{R.}~\bibnamefont{Zandi}},
  \bibinfo{journal}{Phys. Rev. Lett.} \textbf{\bibinfo{volume}{90}},
  \bibinfo{pages}{248101} (\bibinfo{year}{2003}).

\bibitem[{\citenamefont{Zandi et~al.}(2004)\citenamefont{Zandi, Reguera,
  Bruinsma, Gelbart, and Rudnick}}]{ZandiPNAS04}
\bibinfo{author}{\bibfnamefont{R.}~\bibnamefont{Zandi}},
  \bibinfo{author}{\bibfnamefont{D.}~\bibnamefont{Reguera}},
  \bibinfo{author}{\bibfnamefont{R.~F.} \bibnamefont{Bruinsma}},
  \bibinfo{author}{\bibfnamefont{W.~M.} \bibnamefont{Gelbart}},
  \bibnamefont{and} \bibinfo{author}{\bibfnamefont{J.}~\bibnamefont{Rudnick}},
  \bibinfo{journal}{Proc. Nat. Acad. Sci. USA} \textbf{\bibinfo{volume}{101}},
  \bibinfo{pages}{15556} (\bibinfo{year}{2004}).

\bibitem[{\citenamefont{Zandi et~al.}(2006)\citenamefont{Zandi, van~der Schoot,
  Reguera, Kegel, and Reiss}}]{ZandiBJ06}
\bibinfo{author}{\bibfnamefont{R.}~\bibnamefont{Zandi}},
  \bibinfo{author}{\bibfnamefont{P.}~\bibnamefont{van~der Schoot}},
  \bibinfo{author}{\bibfnamefont{D.}~\bibnamefont{Reguera}},
  \bibinfo{author}{\bibfnamefont{W.}~\bibnamefont{Kegel}}, \bibnamefont{and}
  \bibinfo{author}{\bibfnamefont{H.}~\bibnamefont{Reiss}},
  \bibinfo{journal}{Biophys. J.} \textbf{\bibinfo{volume}{90}},
  \bibinfo{pages}{1939} (\bibinfo{year}{2006}).

\bibitem[{\citenamefont{Rapaport}(2008)}]{RapaportPRL08}
\bibinfo{author}{\bibfnamefont{D.~C.} \bibnamefont{Rapaport}},
  \bibinfo{journal}{Phys. Rev. Lett.} \textbf{\bibinfo{volume}{101}},
  \bibinfo{pages}{186101} (\bibinfo{year}{2008}).

\bibitem[{\citenamefont{Roos et~al.}(2010)\citenamefont{Roos, Bruinsma, and
  Wuite}}]{Roos10}
\bibinfo{author}{\bibfnamefont{W.~H.} \bibnamefont{Roos}},
  \bibinfo{author}{\bibfnamefont{R.}~\bibnamefont{Bruinsma}}, \bibnamefont{and}
  \bibinfo{author}{\bibfnamefont{G.~J.~L.} \bibnamefont{Wuite}},
  \bibinfo{journal}{Nature Phys.} \textbf{\bibinfo{volume}{6}},
  \bibinfo{pages}{733} (\bibinfo{year}{2010}).

\bibitem[{\citenamefont{Rapaport}(2010)}]{Rapaport10}
\bibinfo{author}{\bibfnamefont{D.~C.} \bibnamefont{Rapaport}},
  \bibinfo{journal}{J. Phys.: Condens. Matter} \textbf{\bibinfo{volume}{22}},
  \bibinfo{pages}{104115} (\bibinfo{year}{2010}).

\bibitem[{\citenamefont{Klug}(1999)}]{Klug99}
\bibinfo{author}{\bibfnamefont{A.}~\bibnamefont{Klug}}, \bibinfo{journal}{Phil.
  Trans. R. Soc. Lond. B} \textbf{\bibinfo{volume}{354}}, \bibinfo{pages}{531}
  (\bibinfo{year}{1999}).

\bibitem[{\citenamefont{Wertheim}(1984)}]{Wertheim84}
\bibinfo{author}{\bibfnamefont{M.~S.} \bibnamefont{Wertheim}},
  \bibinfo{journal}{J. Stat. Phys.} \textbf{\bibinfo{volume}{35}},
  \bibinfo{pages}{19} (\bibinfo{year}{1984}).

\bibitem[{\citenamefont{Sciortino et~al.}(2007)\citenamefont{Sciortino,
  Bianchi, Douglas, and Tartaglia}}]{SciortinoJCP07}
\bibinfo{author}{\bibfnamefont{F.}~\bibnamefont{Sciortino}},
  \bibinfo{author}{\bibfnamefont{E.}~\bibnamefont{Bianchi}},
  \bibinfo{author}{\bibfnamefont{J.~F.} \bibnamefont{Douglas}},
  \bibnamefont{and}
  \bibinfo{author}{\bibfnamefont{P.}~\bibnamefont{Tartaglia}},
  \bibinfo{journal}{J. Chem. Phys.} \textbf{\bibinfo{volume}{126}},
  \bibinfo{pages}{194903} (\bibinfo{year}{2007}).

\bibitem[{\citenamefont{Dudowicz et~al.}(1999)\citenamefont{Dudowicz, Freed,
  and Douglas}}]{DudowiczJCP99}
\bibinfo{author}{\bibfnamefont{J.}~\bibnamefont{Dudowicz}},
  \bibinfo{author}{\bibfnamefont{K.~F.} \bibnamefont{Freed}}, \bibnamefont{and}
  \bibinfo{author}{\bibfnamefont{J.~F.} \bibnamefont{Douglas}},
  \bibinfo{journal}{J. Chem. Phys.} \textbf{\bibinfo{volume}{111}},
  \bibinfo{pages}{7116} (\bibinfo{year}{1999}).

\bibitem[{\citenamefont{Plimpton}(1995)}]{plimpton95}
\bibinfo{author}{\bibfnamefont{S.~J.} \bibnamefont{Plimpton}},
  \bibinfo{journal}{J. Comp. Phys.} \textbf{\bibinfo{volume}{117}},
  \bibinfo{pages}{1} (\bibinfo{year}{1995}).

\bibitem[{lam()}]{lammps}
\bibinfo{note}{{http://lammps.sandia.gov/}}.

\bibitem[{\citenamefont{Oosawa and Kasai}(1962)}]{OosawaJMB62}
\bibinfo{author}{\bibfnamefont{F.}~\bibnamefont{Oosawa}} \bibnamefont{and}
  \bibinfo{author}{\bibfnamefont{M.}~\bibnamefont{Kasai}}, \bibinfo{journal}{J.
  Mol. Biol.} \textbf{\bibinfo{volume}{4}}, \bibinfo{pages}{10}
  (\bibinfo{year}{1962}).

\bibitem[{\citenamefont{Chr\'{e}tien et~al.}(1995)\citenamefont{Chr\'{e}tien,
  Fuller, and Karsenti}}]{ChretienJCB95}
\bibinfo{author}{\bibfnamefont{D.}~\bibnamefont{Chr\'{e}tien}},
  \bibinfo{author}{\bibfnamefont{S.~D.} \bibnamefont{Fuller}},
  \bibnamefont{and} \bibinfo{author}{\bibfnamefont{E.}~\bibnamefont{Karsenti}},
  \bibinfo{journal}{J. Cell. Biol.} \textbf{\bibinfo{volume}{129}},
  \bibinfo{pages}{1311} (\bibinfo{year}{1995}).

\bibitem[{\citenamefont{Gardner et~al.}(2008)\citenamefont{Gardner, Hunt,
  Goodson, and Odde}}]{GardnerCurrOp08}
\bibinfo{author}{\bibfnamefont{M.~K.} \bibnamefont{Gardner}},
  \bibinfo{author}{\bibfnamefont{A.}~\bibnamefont{Hunt}},
  \bibinfo{author}{\bibfnamefont{H.~V.} \bibnamefont{Goodson}},
  \bibnamefont{and} \bibinfo{author}{\bibfnamefont{D.}~\bibnamefont{Odde}},
  \bibinfo{journal}{Curr. Opin. Cell Biol.} \textbf{\bibinfo{volume}{20}},
  \bibinfo{pages}{64} (\bibinfo{year}{2008}).

\bibitem[{\citenamefont{Kessemakers et~al.}(2006)\citenamefont{Kessemakers,
  Munteanu, laan, Noetzel, Janson, and Dogterom}}]{KerssemakersNature06}
\bibinfo{author}{\bibfnamefont{J.~W.} \bibnamefont{Kessemakers}},
  \bibinfo{author}{\bibfnamefont{E.~L.} \bibnamefont{Munteanu}},
  \bibinfo{author}{\bibfnamefont{L.}~\bibnamefont{laan}},
  \bibinfo{author}{\bibfnamefont{T.~L.} \bibnamefont{Noetzel}},
  \bibinfo{author}{\bibfnamefont{M.~E.} \bibnamefont{Janson}},
  \bibnamefont{and} \bibinfo{author}{\bibfnamefont{M.}~\bibnamefont{Dogterom}},
  \bibinfo{journal}{Nature} \textbf{\bibinfo{volume}{442}},
  \bibinfo{pages}{709} (\bibinfo{year}{2006}).

\bibitem[{\citenamefont{{H. T. Schek~III} et~al.}(2007)\citenamefont{{H. T.
  Schek~III}, Gardner, Cheng, Odde, and Hunt}}]{SchekCB07}
\bibinfo{author}{\bibnamefont{{H. T. Schek~III}}},
  \bibinfo{author}{\bibfnamefont{M.~K.} \bibnamefont{Gardner}},
  \bibinfo{author}{\bibfnamefont{J.}~\bibnamefont{Cheng}},
  \bibinfo{author}{\bibfnamefont{D.~J.} \bibnamefont{Odde}}, \bibnamefont{and}
  \bibinfo{author}{\bibfnamefont{A.~J.} \bibnamefont{Hunt}},
  \bibinfo{journal}{Curr. Biol.} \textbf{\bibinfo{volume}{17}},
  \bibinfo{pages}{1445} (\bibinfo{year}{2007}).

\bibitem[{\citenamefont{Mozziconacci et~al.}(2008)\citenamefont{Mozziconacci,
  Sandblad, Wachsmuth, Brunner, and karsenti}}]{MozziconacciPLoS08}
\bibinfo{author}{\bibfnamefont{J.}~\bibnamefont{Mozziconacci}},
  \bibinfo{author}{\bibfnamefont{L.}~\bibnamefont{Sandblad}},
  \bibinfo{author}{\bibfnamefont{M.}~\bibnamefont{Wachsmuth}},
  \bibinfo{author}{\bibfnamefont{D.}~\bibnamefont{Brunner}}, \bibnamefont{and}
  \bibinfo{author}{\bibfnamefont{E.}~\bibnamefont{karsenti}},
  \bibinfo{journal}{PLoS ONE} \textbf{\bibinfo{volume}{3}},
  \bibinfo{pages}{3821} (\bibinfo{year}{2008}).

\bibitem[{\citenamefont{Erickson}(1989)}]{EricksonJMB89}
\bibinfo{author}{\bibfnamefont{H.~P.} \bibnamefont{Erickson}},
  \bibinfo{journal}{J. Mol. Biol.} \textbf{\bibinfo{volume}{206}},
  \bibinfo{pages}{465} (\bibinfo{year}{1989}).

\bibitem[{\citenamefont{Chen and Hill}(1985)}]{ChenPNAS85}
\bibinfo{author}{\bibfnamefont{Y.~D.} \bibnamefont{Chen}} \bibnamefont{and}
  \bibinfo{author}{\bibfnamefont{T.~L.} \bibnamefont{Hill}},
  \bibinfo{journal}{Proc. Nat. Acad. Sci. USA} \textbf{\bibinfo{volume}{82}},
  \bibinfo{pages}{1131} (\bibinfo{year}{1985}).

\bibitem[{\citenamefont{Bayley et~al.}(1990)\citenamefont{Bayley, Schilstra,
  and Martin}}]{BayleyJCS90}
\bibinfo{author}{\bibfnamefont{P.~M.} \bibnamefont{Bayley}},
  \bibinfo{author}{\bibfnamefont{M.~J.} \bibnamefont{Schilstra}},
  \bibnamefont{and} \bibinfo{author}{\bibfnamefont{S.~R.}
  \bibnamefont{Martin}}, \bibinfo{journal}{J. Cell Sci.}
  \textbf{\bibinfo{volume}{95}}, \bibinfo{pages}{33} (\bibinfo{year}{1990}).

\bibitem[{\citenamefont{Martin et~al.}(1993)\citenamefont{Martin, Schilstra,
  and Bayley}}]{MartinBJ93}
\bibinfo{author}{\bibfnamefont{S.~R.} \bibnamefont{Martin}},
  \bibinfo{author}{\bibfnamefont{M.~J.} \bibnamefont{Schilstra}},
  \bibnamefont{and} \bibinfo{author}{\bibfnamefont{P.~M.}
  \bibnamefont{Bayley}}, \bibinfo{journal}{Biophys. J.}
  \textbf{\bibinfo{volume}{65}}, \bibinfo{pages}{578} (\bibinfo{year}{1993}).

\bibitem[{\citenamefont{Flyvbjerg et~al.}(1996)\citenamefont{Flyvbjerg, Holy,
  and Leibler}}]{FlyvbjergPRE96}
\bibinfo{author}{\bibfnamefont{H.}~\bibnamefont{Flyvbjerg}},
  \bibinfo{author}{\bibfnamefont{T.~E.} \bibnamefont{Holy}}, \bibnamefont{and}
  \bibinfo{author}{\bibfnamefont{S.}~\bibnamefont{Leibler}},
  \bibinfo{journal}{Phys. Rev. E} \textbf{\bibinfo{volume}{54}},
  \bibinfo{pages}{5538} (\bibinfo{year}{1996}).

\bibitem[{\citenamefont{VanBuren et~al.}(2002)\citenamefont{VanBuren, Odde, and
  Cassimeris}}]{VanBurenPNAS02}
\bibinfo{author}{\bibfnamefont{V.}~\bibnamefont{VanBuren}},
  \bibinfo{author}{\bibfnamefont{D.}~\bibnamefont{Odde}}, \bibnamefont{and}
  \bibinfo{author}{\bibfnamefont{L.}~\bibnamefont{Cassimeris}},
  \bibinfo{journal}{Proc. Nat. Acad. Sci. USA} \textbf{\bibinfo{volume}{99}},
  \bibinfo{pages}{6035} (\bibinfo{year}{2002}).

\bibitem[{\citenamefont{Wu et~al.}(2009)\citenamefont{Wu, Wang, Mu, Ouyang,
  Nogales, and Xing}}]{WuPLOS09}
\bibinfo{author}{\bibfnamefont{Z.}~\bibnamefont{Wu}},
  \bibinfo{author}{\bibfnamefont{H.~W.} \bibnamefont{Wang}},
  \bibinfo{author}{\bibfnamefont{W.}~\bibnamefont{Mu}},
  \bibinfo{author}{\bibfnamefont{Z.}~\bibnamefont{Ouyang}},
  \bibinfo{author}{\bibfnamefont{E.}~\bibnamefont{Nogales}}, \bibnamefont{and}
  \bibinfo{author}{\bibfnamefont{J.}~\bibnamefont{Xing}},
  \bibinfo{journal}{PLos ONE} \textbf{\bibinfo{volume}{4}},
  \bibinfo{pages}{e7291} (\bibinfo{year}{2009}).

\bibitem[{\citenamefont{Piette et~al.}(2009)\citenamefont{Piette, Liu, Peeters,
  Smertenko, Hawkins, Deeks, Quinlan, Zakrzewski, and Hussey}}]{PiettePLOS09}
\bibinfo{author}{\bibfnamefont{B.~M. A.~G.} \bibnamefont{Piette}},
  \bibinfo{author}{\bibfnamefont{J.}~\bibnamefont{Liu}},
  \bibinfo{author}{\bibfnamefont{K.}~\bibnamefont{Peeters}},
  \bibinfo{author}{\bibfnamefont{A.}~\bibnamefont{Smertenko}},
  \bibinfo{author}{\bibfnamefont{T.}~\bibnamefont{Hawkins}},
  \bibinfo{author}{\bibfnamefont{M.}~\bibnamefont{Deeks}},
  \bibinfo{author}{\bibfnamefont{R.}~\bibnamefont{Quinlan}},
  \bibinfo{author}{\bibfnamefont{W.~J.} \bibnamefont{Zakrzewski}},
  \bibnamefont{and} \bibinfo{author}{\bibfnamefont{P.~J.}
  \bibnamefont{Hussey}}, \bibinfo{journal}{PLoS ONE}
  \textbf{\bibinfo{volume}{4}}, \bibinfo{pages}{e6378} (\bibinfo{year}{2009}).

\bibitem[{\citenamefont{Mololdtsov et~al.}(2003)\citenamefont{Mololdtsov,
  Ermakova, Shnol, Grishchuk, McIntosh, and Ataullakhanov}}]{MolodtsovBJ05}
\bibinfo{author}{\bibfnamefont{M.~I.} \bibnamefont{Mololdtsov}},
  \bibinfo{author}{\bibfnamefont{E.~A.} \bibnamefont{Ermakova}},
  \bibinfo{author}{\bibfnamefont{E.}~\bibnamefont{Shnol}},
  \bibinfo{author}{\bibfnamefont{E.~L.} \bibnamefont{Grishchuk}},
  \bibinfo{author}{\bibfnamefont{J.~R.} \bibnamefont{McIntosh}},
  \bibnamefont{and} \bibinfo{author}{\bibfnamefont{F.~I.}
  \bibnamefont{Ataullakhanov}}, \bibinfo{journal}{Biophys. J.}
  \textbf{\bibinfo{volume}{88}}, \bibinfo{pages}{3167} (\bibinfo{year}{2003}).

\bibitem[{\citenamefont{VanBuren et~al.}(2005)\citenamefont{VanBuren,
  Cassimeris, and Odde}}]{VanBurenBJ05}
\bibinfo{author}{\bibfnamefont{V.}~\bibnamefont{VanBuren}},
  \bibinfo{author}{\bibfnamefont{L.}~\bibnamefont{Cassimeris}},
  \bibnamefont{and} \bibinfo{author}{\bibfnamefont{D.}~\bibnamefont{Odde}},
  \bibinfo{journal}{Biophys. J.} \textbf{\bibinfo{volume}{89}},
  \bibinfo{pages}{2911} (\bibinfo{year}{2005}).

\end{thebibliography}

\end{document}